\documentclass[%
 reprint,
 amsmath,amssymb,
]{revtex4-1}
\usepackage{graphicx}
\usepackage{dcolumn}
\usepackage{bm}

\begin{document}

\title{Superconductivity and Ferromagnetism in hole-doped RbEuFe$_4$As$_4$}

\author{Yi Liu} \affiliation{Department of Physics, Zhejiang University, Hangzhou 310027, China}

\author{Ya-Bin Liu} \affiliation{Department of Physics, Zhejiang University, Hangzhou 310027, China}

\author{Zhang-Tu Tang} \affiliation{Department of Physics, Zhejiang University, Hangzhou 310027, China}

\author{Hao Jiang}
 \altaffiliation[Present address: ]{School of Physics and Optoelectronics, Xiangtan University, Xiangtan 411105, China}

\author{Zhi-Cheng Wang} \affiliation{Department of Physics, Zhejiang University, Hangzhou 310027, China}

\author{Abuduweli Ablimit} \affiliation{Department of Physics, Zhejiang University, Hangzhou 310027, China}

\author{Wen-He Jiao} \affiliation{Department of Physics, Zhejiang University of Science and Technology, Hangzhou 310023, China}

\author{Qian Tao} \affiliation{Department of Physics, Zhejiang University, Hangzhou 310027, China}

\author{Chun-Mu Feng} \affiliation{Department of Physics, Zhejiang University, Hangzhou 310027, China}

\author{Zhu-An Xu}
\affiliation{Department of Physics, Zhejiang University, Hangzhou
310027, China} \affiliation{State Key Lab of Silicon Materials,
Zhejiang University, Hangzhou 310027, China} \affiliation{Collaborative Innovation Centre of Advanced Microstructures, Nanjing 210093, China}

\author{Guang-Han Cao} \email[]{ghcao@zju.edu.cn}
\affiliation{Department of Physics, Zhejiang University, Hangzhou
310027, China} \affiliation{State Key Lab of Silicon Materials,
Zhejiang University, Hangzhou 310027, China} \affiliation{Collaborative Innovation Centre of Advanced Microstructures, Nanjing 210093, China}

\date{\today}



\begin{abstract}
We discover a robust coexistence of superconductivity and ferromagnetism in an iron arsenide RbEuFe$_4$As$_4$. The new material crystallizes in an intergrowth structure of RbFe$_2$As$_2$ and EuFe$_2$As$_2$, such that the Eu sublattice turns out to be primitive instead of being body-centered in EuFe$_2$As$_2$. The FeAs layers, featured by asymmetric As coordinations, are hole doped due to charge homogenization. Our combined measurements of electrical transport, magnetization and heat capacity unambiguously and consistently indicate bulk superconductivity at 36.5 K in the FeAs layers and ferromagnetism at 15 K in the Eu sublattice. Interestingly, the Eu-spin ferromagnetic ordering belongs to a rare third-order transition, according to the Ehrenfest classification of phase transition. We also identify an additional anomaly at $\sim$ 5 K, which is possibly associated with the interplay between superconductivity and ferromagnetism.

\end{abstract}

\maketitle

\section{\label{sec:level1}Introduction}


Doped EuFe$_2$As$_2$ superconductors stand out among Fe-based superconductors (FeSCs), because the Eu$^{2+}$ local spins ($S$=7/2) may order \emph{ferromagnetically} in the superconducting state\cite{ren2009,felner2011,rxs2014,nd2014}. The undoped EuFe$_2$As$_2$ undergoes an $A$-type antiferromagnetic (namely, in-plane ferromagnetic while inter-plane antiferromagnetic) transition at 19 K in the Eu sublattice\cite{ren2008,jeevan2008,js-njp}, in addition to a spin-density wave (SDW) transition at 190 K in the Fe sublattice. In the magnetically ordered state, the Eu spins align along the crystallographic [110] direction, collinear with the Fe moments\cite{rxs2009,nd2009}. Upon partial P-for-As substitution, which effectively induces a chemical pressure, the Fe-site SDW order is suppressed, and then superconductivity (SC) emerges at $T_{\mathrm{sc}}\sim$ 26 K\cite{ren2009,felner2011,cao2011,jeevan2011}. Simultaneously, the Eu spins become ferromagnetically ordered at $T_{\mathrm{Curie}}\sim$ 20 K, accompanying with a spin reorientation towards the [001] (or $c$-axis) direction\cite{felner2011}. Similar coexistence of Fe-based SC and Eu-spin ferromagnetism (FM) was also observed and verified in several Fe-site electron-doped EuFe$_2$As$_2$ systems\cite{jiang2009,jiao2011,jiao2013,jin-Co,jin-Ir}.

Nevertheless, there have been some debates on the details of the Eu-spin ordering. In Refs.~\cite{jeevan2011,tokiwa}, it was argued that SC coexists with Eu-spin antiferromagnetism (AFM) in the superconducting regime of EuFe$_2$(As$_{1-x}$P$_x$)$_2$. Later, Zapf et al.\cite{zapf2011} proposed that SC coexists with a canted AFM, such that it is virtually ferromagnetic along the $c$ axis. These authors\cite{zapf2013} further revised the electronic phase diagram because of the discovery of a reentrant spin glass state. Recent x-ray resonant magnetic scattering\cite{rxs2014} and neutron scattering\cite{nd2014} experiments however indicated long-range ferromagnetic orderings for Eu spins in superconducting EuFe$_2$(As$_{1-x}$P$_x$)$_2$ with $x$ = 0.19 and 0.15, respectively. It was demonstrated that the Eu spins align exactly along the $c$ axis, in contradiction to the spin-canting scenario. So far, this discrepancy remains unresolved. Note that the spin-tilting angle ($\sim$20$^{\circ}$ from the $c$ axis, as detected by M\"{o}ssbauer measurements\cite{felner2011}) coincides with the direction that connects the interlayer next-nearest (NN) Eu atoms because of the body-centered Eu sublattice. To clarify whether the Eu-sublattice type is relevant to Eu spin orientations, it is desirable to study a related material system in which Eu atoms form a primitive tetragonal lattice.

Local-moment FM and spin-singlet SC are known to be mutually incompatible\cite{ginzburg,rev1985,maple}, which makes their coexistence (hereafter abbreviated as FM+SC) very rare\cite{fmsc}. The FM+SC phenomenon observed in FeSCs has been ascribed to the multi-orbital character as well as the robustness of superconductivity against magnetic fields\cite{cao2011,cao2012}. On the one hand, the zero-temperature upper critical magnetic field, $H_{\mathrm{c2}}(0)$, of FeSCs is typically higher than 50 T\cite{johnston,stewart}, which is large enough to fight the internal exchange field that is comparable to the hyperfine field on the Eu nucleus ($\sim$ 25 T)\cite{felner2011}. On the other hand, the Eu-spin FM can be satisfied even in the presence of SC, because the Fe-3$d$ multi-orbitals enable both superconducting pairing (dominated by the $d_{yz}$ and $d_{zx}$ electrons\cite{zhangsc}) and the Ruderman-Kittel-Kasuya-Yosida (RKKY) exchange interaction between Eu local moments. The RKKY interaction can be mediated by different Fe-3$d$ orbitals such as $d_{x^{2}-y^{2}}$ and $d_{z^{2}}$\cite{ren-Ni}. Therefore, both SC and local-moment FM can be favored in FeSCs\cite{cao2012}.

The crucial factor that leads to Eu local-moment FM should be the interlayer RKKY interaction, since the in-plane RKKY coupling remains ferromagnetic even in the parent compound EuFe$_2$As$_2$. The interlayer exchange coupling ($J_{\mathrm{R}}^{\perp}$) is simplified to be proportional to $\mathrm{cos}(2k_{\mathrm{F}}r)/r^{3}$ for a large $r$, where $k_\mathrm{F}$ is the Fermi wave vector and $r$ denotes the distance between local moments. This means that $J_{\mathrm{R}}^{\perp}$ can be changed from negative (AFM) to positive (FM) by tuning the $k_{\mathrm{F}}r$ value. The AFM-to-FM transition in EuFe$_2$(As$_{1-x}$P$_x$)$_2$\cite{cao2011} and Eu(Fe$_{1-x}$Ni$_x$)$_2$As$_2$\cite{ren-Ni} is mainly due to the change in $k_\mathrm{F}$ (where the heavy three-dimensional hole pocket\cite{singh} seems to be involved), simply because the interlayer Eu interatomic distance ($r_{\perp}$) varies only slightly. Indeed, based on a first-principles calculation\cite{LDA}, the effective interlayer NN magnetic coupling changes from antiferromagnetic in EuFe$_2$As$_2$ to ferromagnetic in EuFe$_2$P$_2$, while the in-plane coupling remains ferromagnetic.

The above statement suggests an alternative approach to realize the AFM-to-FM transition which may lead to an FM+SC state as well. By constructing a crystal structure in which $r_{\perp}$ changes significantly, the sign of $J_{\mathrm{R}}^{\perp}$ may be altered accordingly. We previously designed a related structure, exemplified as KLaFe$_4$As$_4$ (1144)\cite{jh}, which can be viewed as an intergrowth of KFe$_2$As$_2$ and LaFe$_2$As$_2$. In the Eu analog, $Ak$EuFe$_4$As$_4$ ($Ak$ denotes an alkali metal), the Eu atoms form a \emph{primitive} tetragonal lattice, such that the lines that connect the interlayer-NN Eu atoms are exactly parallel to the $c$ axis. Notably, the $r_{\perp}$ value is roughly doubled because every alternate Eu atoms are replaced by $Ak$ along the $c$ axis.

Very recently, we became aware of the report of $AkAe$Fe$_4$As$_4$ ($Ae$=Ca, Sr, Ba) superconductors which possess the identical 1144-type structure\cite{1144}. This work inspires us to re-investigate our target material $Ak$EuFe$_4$As$_4$. Consequently, we succeeded in synthesizing a 1144 compound RbEuFe$_4$As$_4$. In this paper, we report the crystal structure and physical properties of this new material. We indeed observe an Eu-spin FM at 15 K which coexists with a bulk SC at an unexpectedly high $T_{\mathrm{sc}}$ of 36.5 K. Remarkably, the evidence for FM+SC are very robust, compared to the previous FM+SC phenomena in doped EuFe$_2$As$_2$ systems\cite{ren2009,jeevan2011,jiao2011,jiao2013}. Apart from FM+SC, additional intriguing phenomena were also observed and discussed.

\section{\label{sec:level2}Experimental Methods}

RbEuFe$_4$As$_4$ polycrystalline samples were prepared via a solid-state reaction method. First, FeAs, EuAs and RbAs were prepared using source materials Rb (99.75\%), As (99.999\%) and Eu (99.9\%) pieces and Fe powder (99.998\%). The presynthesized arsenides were then ball milled separately for 10 minutes in a glove box filled with pure Ar (the water and oxygen content is below 1 ppm). Second, stoichiometric mixtures (namely, Rb:Eu:Fe:As=1:1:4:4) of Fe, FeAs, EuAs and RbAs were homogenized, pressed into pellets, and then loaded in an alumina tube-like crucible which was sealed in a Ta tube. The Ta tube was protected in a quartz ampoule filled with Ar gas ($\sim$ 0.6 bar). Third, the sample-loaded ampoule was heated rapidly to 1123-1173 K in a muffle furnace. After holding for 6 hours, the sample was quenched, similar to the approach in synthesizing $AkAe$Fe$_4$As$_4$\cite{1144}. The sample's quality could be improved by repeating the solid-state reaction.

Powder x-ray diffraction (XRD) was carried out on a PANalytical x-ray diffractometer with a monochromatic Cu-K$_{\alpha1}$  radiation. The crystal structure was refined by a Rietveld analysis using a RIETAN software\cite{rietan-fp2}. The electrical transport and heat capacity measurements were conducted on a Quantum Design physical property measurement system (PPMS-9). In the resistivity measurement, a four-electrode method and the ac transport option were employed. The sample pellet was cut into a thin rectangular bar with a dimension of 2.2$\times$1.1$\times$0.5 mm$^3$, on which thin gold wires were attached with silver paint. The excitation current was set to 5.18 mA. The Hall coefficient was measured by permutating the voltage and current electrodes\cite{sample1987}, using a thin-square sample (1.3$\times$1.3$\times$0.12 mm$^3$) with four symmetric electrodes attached. The excitation current was 20 mA, and the applied magnetic field was 80 kOe. The heat capacity was measured by a thermal relaxation method using a square-shaped sample plate with a mass of 19.5 mg. The dc magnetization for a regular shape sample (in order to estimate the demagnetization factor) was measured on a Quantum Design MPMS-5 equipment. Different kind of protocols of zero-field cooling (ZFC) and field-cooling (FC) were employed for probing the superconducting and magnetic transitions.

\section{\label{sec:level3}Results and discussion}

\subsection{\label{subsec:level1}Crystal structure}

The as-prepared sample was characterized by powder XRD. The result shows that most of the reflections can be indexed using a tetragonal lattice, whose unit-cell size ($a\sim$ 3.89 {\AA} and $c\sim$ 13.31 {\AA}) is close to other 1144-type compounds\cite{1144}. The intensity of the strongest reflection of the impurity phase, which is identified to be unreacted FeAs, is only 2.5\% of that of the (103) diffraction peak of the main phase, indicating high quality of the sample. The appearance of ($hkl$) reflections with $h+k+l$=odd numbers confirms that the tetragonal lattice is primitive rather than body centered.

\begin{figure}
\includegraphics[width=8cm]{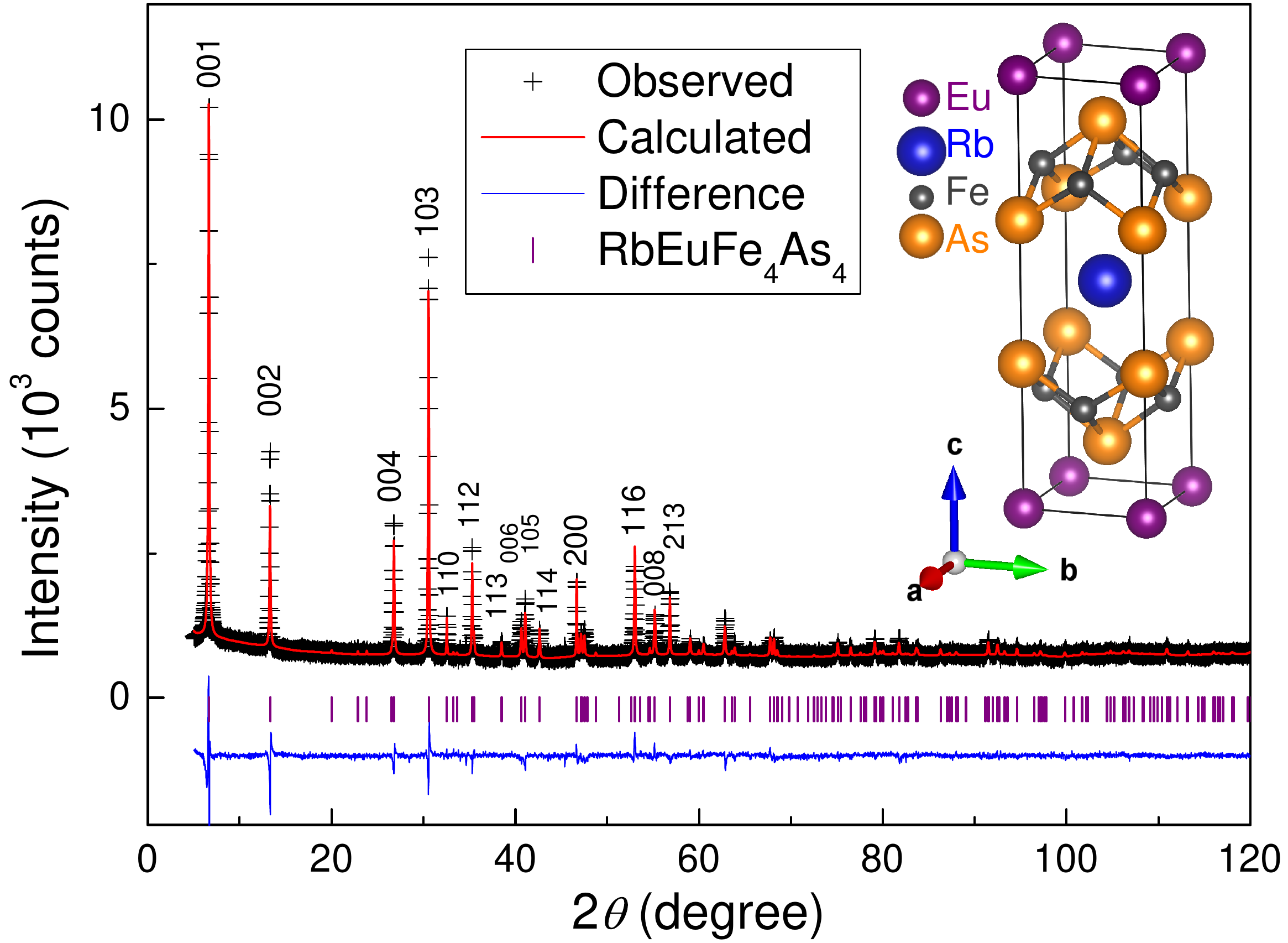}
\label{fig1} \caption{Rietveld refinement profile for the powder x-ray diffraction of RbEuFe$_4$As$_4$ whose crystal structure is displayed in the inset. The relatively strong reflections in low diffraction angles are indexed.}
\end{figure}

Figure 1 shows the Rietveld refinement profile for the powder XRD data of RbEuFe$_4$As$_4$. The refinement yields a weighted reliable factor $R_{\text{wp}}$ of 3.45\% and a ``goodness-of-fit" parameter $S$ of 1.4, which indicate high reliability for the crystal structure refined. As shown in the inset of Fig. 1, it is indeed an intergrowth of RbFe$_2$As$_2$ and EuFe$_2$As$_2$. It can be also viewed as a modified EuFe$_2$As$_2$ in which every alternate Eu atoms are replaced by Rb atoms along the $c$ axis. It is the structural modification that changes the lattice from body-centered to primitive. Additionally, the local coordination environment for Fe atoms turns out to be asymmetric. According to the refined structural parameters listed in Table~\ref{structure}, the Fe$-$As bondlengths are no longer equal. Consequently, As1 (at Rb side) and As2 (at Eu side) heights from the Fe plane are remarkably different (1.249 {\AA} and 1.467 {\AA}, respectively). Meanwhile, the bond angles As1$-$Fe$-$As1 and As2$-$Fe$-$As2 are 112.50$^{\circ}$ and 109.04$^{\circ}$, respectively. Note that the As relative heights as well as the bond angles are opposite to the result in RbCaFe$_4$As$_4$\cite{1144}. This suggests that the Fe-coordination asymmetry does not interfere with the occurrence of SC, although it could influence the superconducting pairing symmetry.
\begin{table}
\caption{\label{structure}
Crystallographic data of RbEuFe$_4$As$_4$ at room temperature with $a$ = 3.8897(1) \r{A}, $c$ = 13.3146(6) \r{A} and space group $P4/mmm$ (No. 123).}
\begin{ruledtabular}
\begin{tabular}{cccccc}
Atom& Wyckoff& $x$ &$y$&$z$&$B (\mathrm{A^{-2}})$ \\
\hline
Eu & 1$a$&0& 0  &0  &0.7(2)\\
Rb & 1$d$&0.5& 0.5 &0.5  &1.8(3) \\
Fe & 4$i$&0&0.5  &0.2404(6) &1.3(2)\\
As1 &2$g$&0&  0 &0.3342(5) &1.2 (2)\\
As2 &2$h$&0.5&0.5   &0.1300(5) &0.7(2)\\
\end{tabular}
\end{ruledtabular}
\end{table}
Comparison of the lattice parameters of RbEuFe$_4$As$_4$ with those of EuFe$_2$As$_2$ and RbFe$_2$As$_2$ hints the ``interaction" between the two building blocks. The $a$ axis is 0.009 (2) {\AA} smaller than the average of those of EuFe$_2$As$_2$\cite{ren2008} and RbFe$_2$As$_2$\cite{Rb122}, meanwhile, the $c$ axis is 0.020(5) {\AA} smaller than half of the sum of those of EuFe$_2$As$_2$ and RbFe$_2$As$_2$. This lattice shrinkage suggests stabilization of the hybrid 1144 phase. The interaction of the two building units is also manifested by the shortening of the ``RbFe$_2$As$_2$" block (from 7.267 {\AA} to 6.914 {\AA}) and the stretching of the ``EuFe$_2$As$_2$" block (from 6.068 {\AA} to 6.401 {\AA}). This structural variation seems to be associated with the charge redistribution, since the Fe formal valence in RbEuFe$_4$As$_4$ has to be averaged to 2.25+ instead of being either 2+ in EuFe$_2$As$_2$ or 2.5+ in RbFe$_2$As$_2$. The elongation of the ``EuFe$_2$As$_2$" block suggests weakening of the effective coupling between Eu-4$f$ spins and Fe-3$d$ itinerant electrons in  RbEuFe$_4$As$_4$.
\subsection{\label{subsec:level2}Electrical resistivity}

\begin{figure}
\includegraphics[width=8cm]{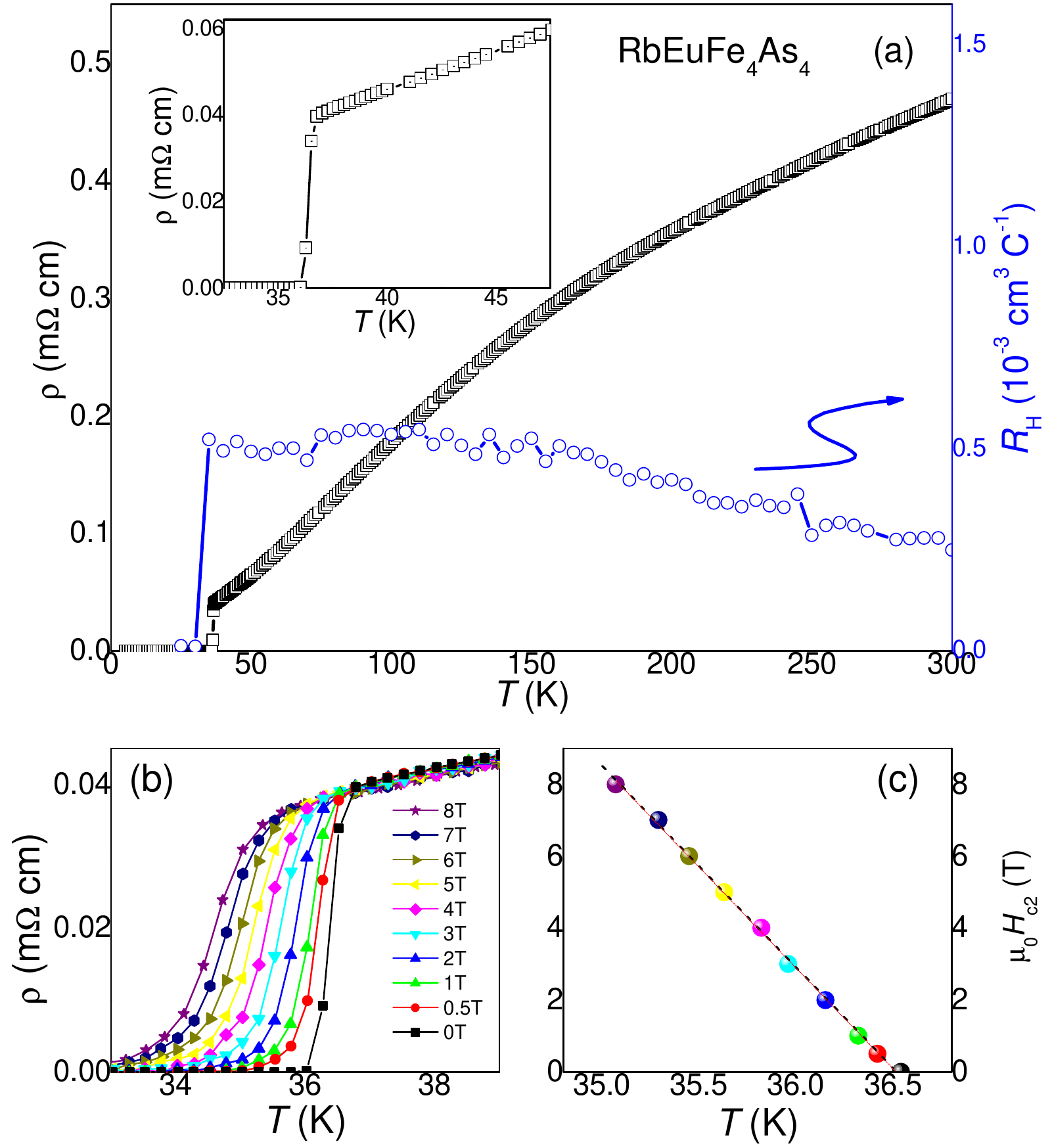}
\label{fig2} \caption{(a) Temperature dependence
of resistivity (left axis) and Hall coefficient (right axis) of the RbEuFe$_4$As$_4$ polycrystalline sample. The inset zooms in the superconducting transition under zero field. (b) The superconducting resistive transitions under increased external magnetic fields. (c) Plot of the upper critical field as a function of temperature, derived from the data shown in (b). The dashed line is the linear fit.}
\end{figure}
Figure 2 shows electrical transport measurement results for the RbEuFe$_4$As$_4$ polycrystalline sample. The resistivity $\rho(T)$ exhibits a metallic behavior with a broad hump around 180 K, a common feature of hole-doped FeSCs\cite{BaK,whh-transport,1144}. As expected, the conduction is dominated by a hole-carrier transport, which is verified by the positive Hall coefficient ($R_{\mathrm{H}}$) in the whole temperature range of the normal state. The $R_{\mathrm{H}}$ values appear to be extremely low (equivalent to a hole content of $n\sim$1.0 holes/Fe using the simple formula $n=1/(eR_{\mathrm{H}}$) for a single-band case), which indicates a multiband scenario (the electron-hole compensation effect accounts for the low $R_{\mathrm{H}}$)\cite{whh-transport}. A sharp superconducting transition appears at 36.5 K, and zero resistance is achieved at 36 K (note that the transition temperature in $R_{\mathrm{H}}(T)$ is decreased because it was measured under an 80-kOe magnetic field). This $T_{\mathrm{sc}}$ value is the highest among all the doped EuFe$_2$As$_2$ superconductors\cite{EuK,ren2009,jiao2011,jiao2013}. Furthermore, no re-entrance superconductivity can be observed, in contrast to the case in the EuFe$_2$As$_2$-related ferromagnetic superconductors where non-zero resistance often re-enters the superconducting state\cite{ren2009,jiang2009,jiao2011,jiao2013}.
Under external magnetic fields, the superconducting transitions shift mildly to lower temperatures, as shown in Fig. 2(b). From these data, the upper critical fields, $H_{\mathrm{c2}}(T)$, can be obtained by defining the transition temperature $T_{\mathrm{sc}}(H)$ at which the resistivity drops to 90\% of the linearly extrapolated one (this criterion satisfies the result of heat-capacity measurement). The resulting $H_{\mathrm{c2}}(T)$, shown in Fig. 2(c), is almost linear, in contrast with the pronounced positive curvature observed in EuFe$_2$(As$_{0.7}$P$_{0.3}$)$_2$\cite{ren2009} and Eu(Fe$_{0.88}$Ir$_{0.12}$)$_2$As$_2$\cite{jiao2013}. The initial slope, $\mu_0$d$H_{\mathrm{c2}}$/d$T|_{T_{\mathrm{sc}}}$, is as large as $-$5.6 T/K, over 4 times of that of EuFe$_2$(As$_{0.7}$P$_{0.3}$)$_2$. Here we note that the measurement was performed on the polycrystalline sample, the $H_{\text{c2}}(T)$ data obtained actually represent some kind of average of $H_{\text{c2}}^{\parallel c}(T)$ and $H_{\text{c2}}^{\parallel ab}(T)$.


\subsection{\label{subsec:level3}Magnetic properties}

Figure 3 shows the dc magnetic susceptibility measured under low magnetic fields using FC and ZFC protocols, respectively. Consistent with the resistivity measurement above, a superconducting onset transition occurs at $T_{\text{sc}}^{\mathrm{onset}}$ =36.5 K. The volume fraction of magnetic shielding, measured in a ZFC process, is almost 100\% at the lowest temperature after making a demagnetization correction. The volume fraction of magnetic repulsion (also called Meissner volume fraction), scaled by the drop in $4\pi \chi_{\mathrm{FC}}$ in a FC process, is reduced to 16\% due to flux pinning while cooling down under magnetic fields. Nevertheless, the Meissner fraction is still over six-fold higher than any impurity content (estimated to be less than 2.5\% from the XRD) in the sample. Therefore, bulk superconductivity is clearly demonstrated for RbEuFe$_4$As$_4$. In comparison, the previous related superconductor EuFe$_2$(As$_{0.7}$P$_{0.3}$)$_2$ shows no diamagnetism even in a ZFC process for the polycrystalline sample\cite{ren2009}. This facts highlight the robustness of SC in the present RbEuFe$_4$As$_4$ system.

\begin{figure}
\includegraphics[width=8cm]{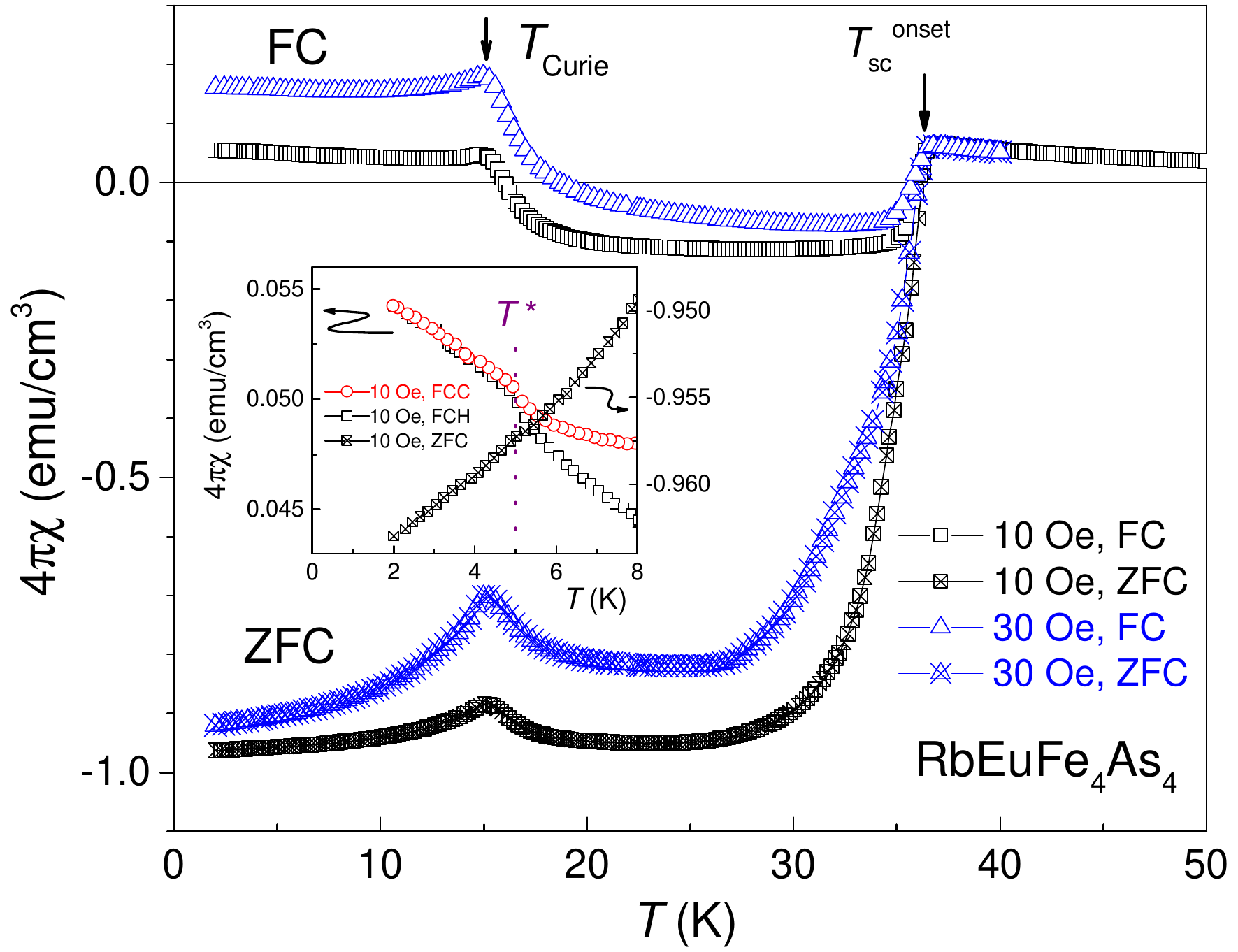}
\label{fig3} \caption{Magnetic susceptibility of the RbEuFe$_4$As$_4$ polycrystalline sample under low magnetic fields in field-cooling (FC) and zero-field-cooling (ZFC) modes, respectively. The superconducting transition at $T_{\text{sc}}^{\mathrm{onset}}$ = 36.5 K and the magnetic transition at $T_{\text{Curie}}$ = 15 K are marked by arrows, respectively. The inset magnifies low-temperature data from which a minor anomaly at $T^{*}$ = 5 K can be distinguished. Right axis: FC data measured in cooling (FCC) and heating (FCH) processes; left axis: ZFC data.}
\end{figure}

Interestingly, there is an anomaly at around 15 K in the superconducting state, featured by a sign change in $\chi_{\mathrm{FC}}$ at $H$=10 Oe. At first sight, this behavior resembles the ac magnetic susceptibility ($\chi_{\mathrm{ac}}$) curve of the classical reentrant superconductor ErRh$_4$B$_4$ in which SC \emph{disappears} when the Er magnetic long-range order sets in\cite{ErRh4B4}. As a matter of fact, however, $\chi_{\mathrm{ac}}$ is equivalent to $\chi_{\mathrm{ZFC}}$ rather than $\chi_{\mathrm{FC}}$. The $\chi_{\mathrm{ZFC}}$ value of RbEuFe$_4$As$_4$ remains diamagnetic (in contrast with the positive $\chi_{\mathrm{ac}}$ in ErRh$_4$B$_4$) in the region where $\chi_{\mathrm{FC}}$ becomes positive, indicating SC survives at the low temperatures. When the applied field is increased to 30 Oe, the $\chi_{\mathrm{FC}}$ value is enhanced remarkably, suggesting that it is possibly associated with a ferromagnetism rather than a paramagnetism. Additionally, there is a minor kink at $T^{*}\sim$ 5 K in both FC and ZFC data, which can be seen in the magnified plot shown in the inset of Fig. 3. The bifurcation of FCC (measured in cooling process) and FCH (measured in heating process) can also be seen at $T>$ 5 K. We will discuss the possible implications later on.

In order to further identify the 15-K transition under the superconducting state, we performed isothermal magnetization [$M(H)$] measurements. As shown in Fig. 4, the $M(H)$ data are essentially linear at $T>T_{\text{sc}}$, indicating a simple paramagnetic state. When the temperature is decreased to 30 K, which is below $T_{\text{sc}}$ but well above 15 K, a typical superconducting loop is superposed on the paramagnetic background. At $T$ = 20 K, which is close to the transition temperature, the paramagnetic component of the $M(H)$ curve behaves in a shape of a Brillouin function. Below 15 K, the overall $M(H)$ loops look like a ferromagnet, but the magnetization curves do not merge together at fields higher than the saturation one (about 1.5 kOe). This is due to the existence of SC which shows a flux-pinning effect as mentioned above. Note that the saturation magnetization at 2 K achieves 6.5 $\mu_{\text{B}}$/Eu, basically consistent with the expected value of $gS$=7.0 $\mu_{\text{B}}$/Eu for the Eu$^{2+}$-spin ferromagnetic alignment. The above results undoubtfully indicate an Eu-spin FM at $T_{\text{Curie}}$ = 15 K for RbEuFe$_4$As$_4$, although the spin orientation is not clear. The robust ferromagnetic properties contrast with those of the previous ferromagnetic superconductor EuFe$_2$(As$_{0.7}$P$_{0.3}$)$_2$ which shows a much lower coersive field (20 Oe at 2 K)\cite{ren2009} and a much higher saturation field ($\sim$ 7 kOe)\cite{cao2011}.

\begin{figure}
\includegraphics[width=8cm]{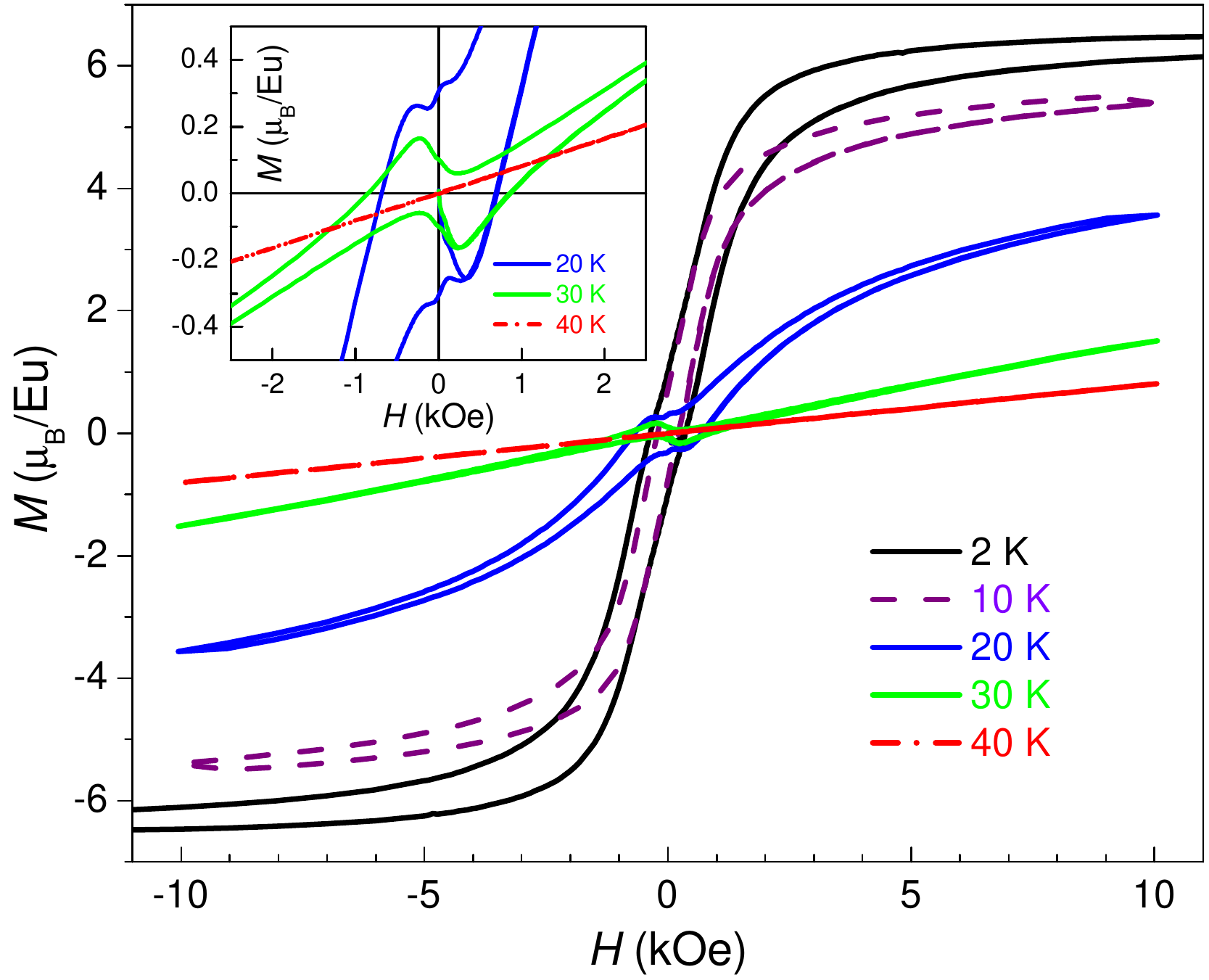}
\label{fig4} \caption{Isothermal magnetization of RbEuFe$_4$As$_4$ at several representative temperatures. The top-left inset zooms in the data in the low-field region for $T$ = 20, 30 and 40 K.}
\end{figure}

Figure 5 plots the magnetic susceptibility $M/H$ (left axis) and its reciprocal (right axis) of RbEuFe$_4$As$_4$ as functions of temperature. The $M/H$ data exhibit a linear dependence in the high-$T$ region, indicating dominant Curie-Weiss paramagnetism. We thus fit the data from 50 to 300 K by an extended Curie's law, $\chi = \chi_0 + C/(T - \theta)$, which yields $\chi_0$ = 0.00178 emu mol$^{-1}$, $C$ = 7.91 emu K mol$^{-1}$ and $\theta$ = 23.6 K. If one assumes that $\chi_0$ is mostly contributed from Pauli paramagnetism, a density of state at Fermi level [$N(E_\mathrm{F})$] can be estimated to be 55 eV$^{-1}$ fu$^{-1}$, which is unusually large. From the Curie constant $C$, an effective local moment of $\mu_{\text{eff}}$ = 7.95 $\mu_{\text{B}}$ fu$^{-1}$ (fu denotes formula unit) is obtained, which is almost equal to the expected value of $g\sqrt{S(S+1)}=$7.94 ($\mu_{\text{B}}$) for an Eu$^{2+}$ spin. The paramagnetic Curie temperature $\theta$ is positive, indicating dominant ferromagnetic interactions among the Eu spins. Indeed, in the low-$T$ data shown in the inset of Fig. 5, a canonical ferromagnetic transition can be seen. The transition temperature can be estimated by the dip in d$M$/d$T$\cite{jiang2009}, which is about 2 K higher than the $T_{\text{Curie}}$ value at $H$ = 10 Oe.

\begin{figure}
\includegraphics[width=8cm]{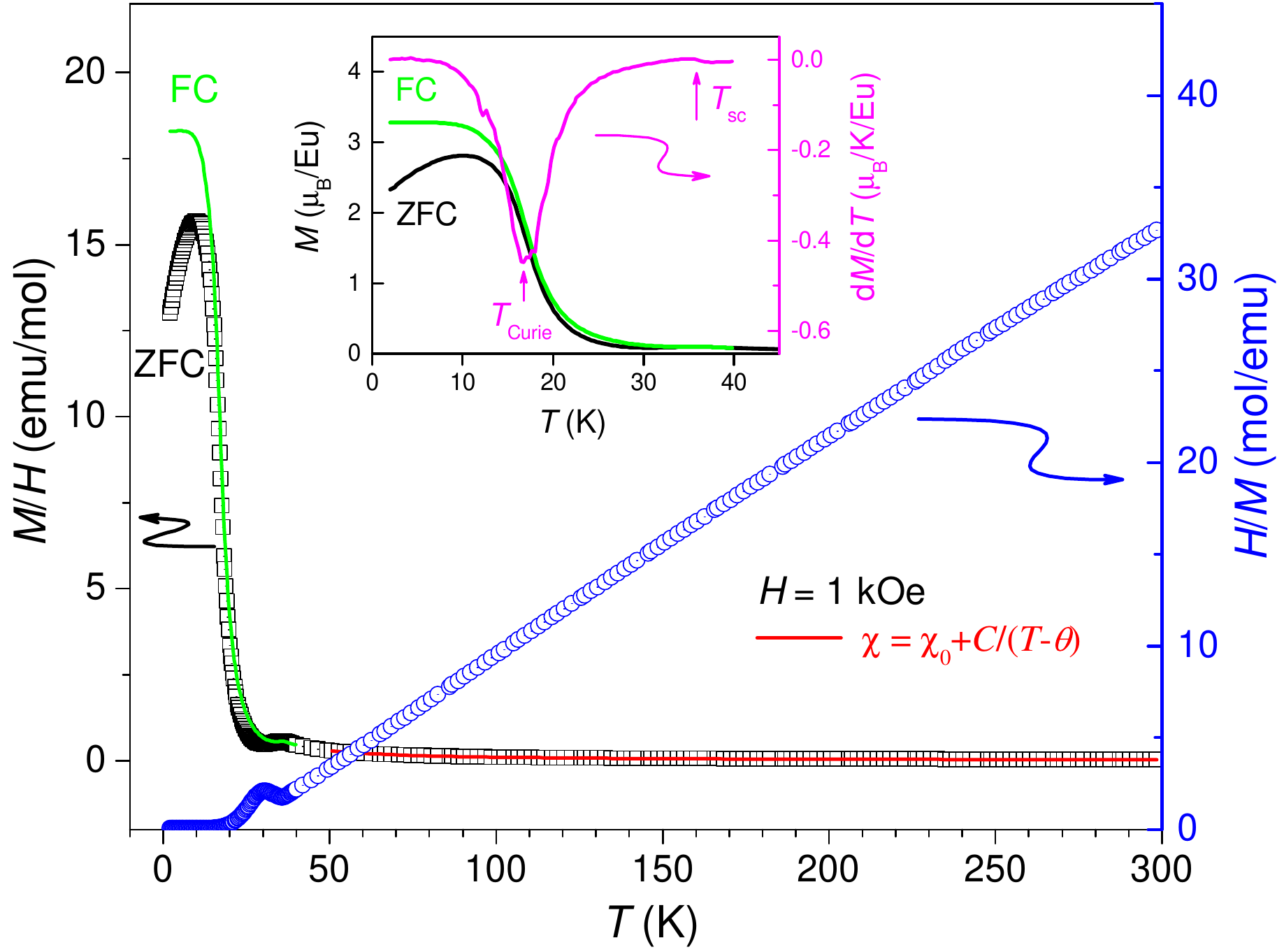}
\label{fig5} \caption{Magnetic susceptibility ($M/H$) of RbEuFe$_4$As$_4$ measured under $H$ = 1 kOe. The data from 50 to 300 K are fitted by the extended Curie's law. The reciprocal of susceptibility ($H/M$) is shown using right axis. The inset shows temperature dependence of magnetization in the low-$T$ region, in which a canonical ferromagnetic transition can be seen. The $T_{\text{Curie}}$ value can be estimated from the derivative of magnetization (right axis).}
\end{figure}

\subsection{\label{subsec:level4}Specific heat}

To further characterize the superconducting and magnetic transitions, we conducted heat capacity measurement for RbEuFe$_4$As$_4$. From the raw $C(T)$ data shown in Fig. 6(a), two anomalies at $T_{\text{Curie}}$ = 15 K and $T_{\text{sc}}$ = 36.5 K can be identified, respectively, verifying the ferromagnetic and superconducting transitions demonstrated above. The inset shows the superconducting transition more clearly. The thermodynamic transition temperature, determined by entropy-conserving construction, is 36.0 K, coincident with the zero-resistance temperature. Impressively, the specific-heat jump $\Delta C$ is as high as 7.5 J K$^{-1}$ mol$^{-1}$, which further confirms bulk SC in RbEuFe$_4$As$_4$. The $\Delta C/T_{\mathrm{sc}}$ value achieves 208 mJ K$^{-2}$ mol$^{-1}$. Thus the electronic specific-heat coefficient $\gamma$ can be estimated to be 145 mJ K$^{-2}$ mol$^{-1}$ by assuming $\Delta C/(\gamma T_{\mathrm{sc}})$ = 1.43 in the BCS weak-coupling scenario. A similarly large $\gamma$ value of $\sim$ 150 mJ K$^{-2}$ mol$^{-1}$ can be \emph{independently} estimated by the enhanced room-temperature specific heat that is 46 J K$^{-1}$ mol$^{-1}$ larger than the Dulong-Petit limit 3$NR$ ($N$ = 10, being the number of atoms in a formula unit). Therefore, the real $\gamma$ value should be around 150 mJ K$^{-2}$ mol$^{-1}$, equivalent to 38 mJ K$^{-2}$ mol-Fe$^{-1}$. Such an enhanced $\gamma$ is often observed in hole-doped FeSCs\cite{whh2009}. Notably, the estimated $\gamma$ value corresponds to $N(E_\mathrm{F})\approx$ 60 eV$^{-1}$ fu$^{-1}$, consistent with the value derived from the magnetic measurement. Namely, the Wilson ratio is about unity albeit of an enhanced $N(E_\mathrm{F})$.


\begin{figure*}
\includegraphics[width=16cm]{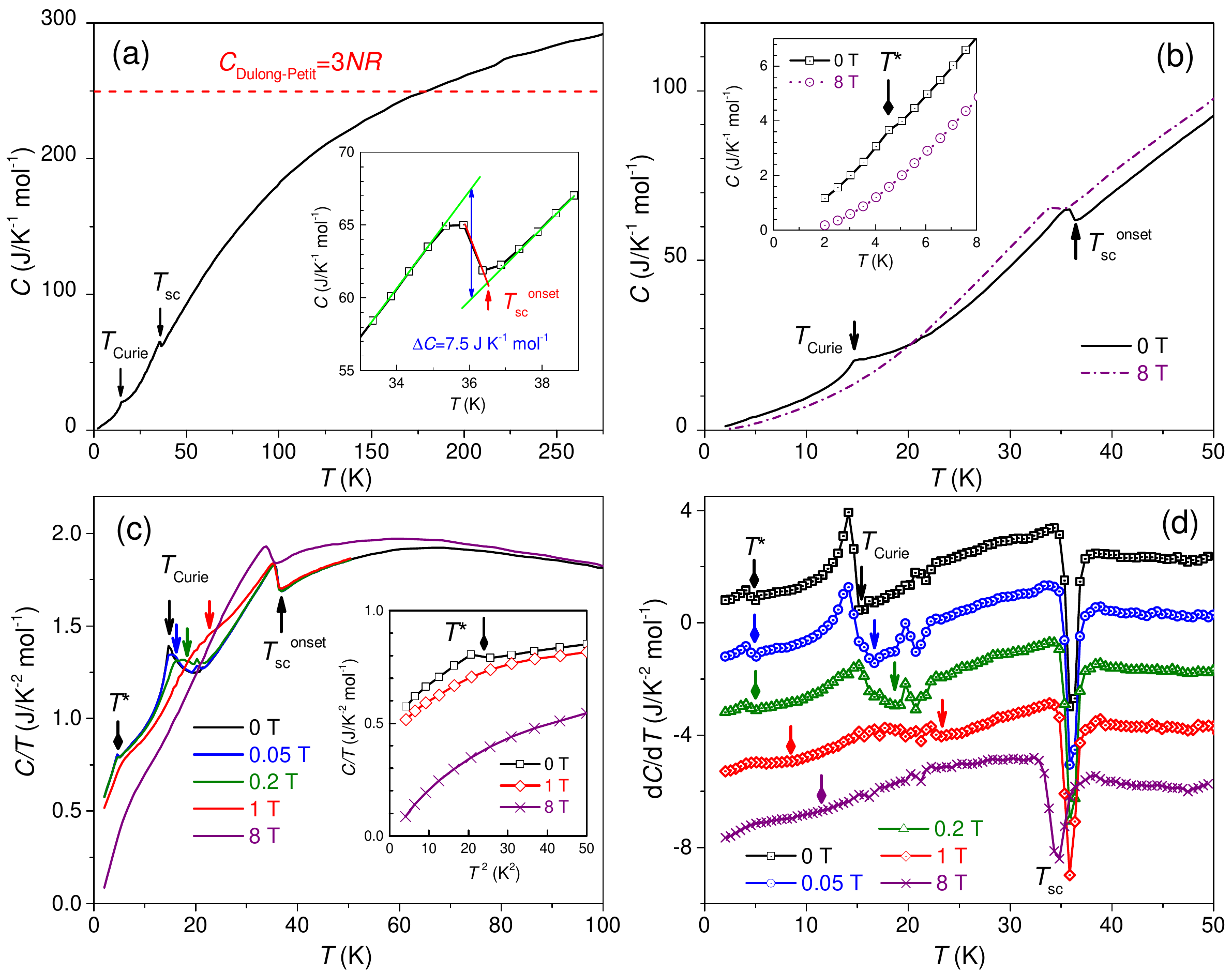}
\label{fig6} \caption{Heat capacity measurements for RbEuFe$_4$As$_4$. The characteristic temperatures of superconducting transition ($T_{\text{sc}}$), ferromagnetic transition ($T_{\text{Curie}}$) and an unknown possible transition ($T^{*}$) are indicated by arrows, respectively. (a) Raw data of the specific heat, $C(T)$, at zero field. The inset zooms in the superconducting transition where a large specific-heat jump is determined. Panel (b) compares the $C(T)$ data at zero field and at 8 T. The inset magnifies the plot below 8 K, from which an anomaly can be distinguished at zero field. Panel(c) plots $C/T$ vs $T$ under various magnetic fields. The inserted plot shows $C/T$ vs $T^2$ for the low-$T$ data. Panel (d) shows the derivative of specific heat (the data under fields are shifted downward for clarity).}
\end{figure*}

Figure 6(b) shows the variations in $C(T)$ under a magnetic field of 8 T. First, the superconducting transition temperature decreases slightly to 35.4 K, consistent with the result of magneto-resistivity measurement above. Second, the anomaly at $T_{\mathrm{Curie}}$ is smeared out by the field. In addition, a minor anomaly at $T^{*}\sim$ 5 K, which can be seen more easily in Fig. 6(c) and (d), tends to disappear as well. Third, the external field induces a $C(T)$-weight transfer from lower to higher temperatures. At $T<$ 20 K, $C(T)$ is suppressed by the field; while in the temperature range of 22 K $<T<$ 100 K, $C(T)$ is enhanced substantially. The field-induced change in $C(T)$ has to be ascribed to the Eu-spin magnetism, because electronic and phonon specific heat ($C_{\mathrm{el}}$ and $C_{\mathrm{L}}$) generally do not change with magnetic fields. The external fields force the Eu spins along the field direction, which severely broadens the phase transition, making the $C(T)$ weight shifts to higher temperatures consequently. Similar effect is also observed in EuFe$_2$(As$_{0.7}$P$_{0.3}$)$_2$\cite{ren2009}.

The $C(T)$-weight transfer under different magnetic fields is explicitly shown in the $C/T$ plot shown in Fig. 6(c). Besides, an upward shift of $T_{\text{Curie}}$ can also be seen, further supporting the ferromagnetic transition. At $T^{*}\sim$ 5 K, the zero-field $C/T$ curve exhibits a tiny jump, which also changes with magnetic fields. Under a high field, the jump turns into a shoulder that moves to high temperatures with increasing field. This sensitive response to magnetic fields suggests that it should be an intrinsic phenomenon. We will return to this topic in the Discussion section.

Shown in the inset of Fig. 6(c) is the $C/T$ vs $T^2$ plot for the low-$T$ data. As is known, the expected $C/T$ value at $T\rightarrow$ 0 K is zero for a fully-gaped superconductor, which increases linearly with magnetic fields. However, the situation here is apparently abnormal: the zero-field $C/T$ value is as large as 573 mJ K$^{-2}$ mol$^{-1}$ at 2 K, which is then suppressed to 86 mJ K$^{-2}$ mol$^{-1}$ at 8 T. This result suggests that the Eu-spin magnetic contribution ($C_{\mathrm{Eu}}$) remains dominant around 2 K. Future measurements down to lower temperatures are expected to reveal the Eu-spin wave excitations as well as the quasi-particle excitations from the superconducting state.

Intriguingly, the specific-heat anomaly at $T_{\text{Curie}}$ is very different from those of most magnetic orderings (such as those in EuFe$_2$As$_2$\cite{ren2008} and EuFe$_2$(As$_{0.7}$P$_{0.3}$)$_2$\cite{ren2009}) which show a clear jump because of a second-order transition. In the present case for RbEuFe$_4$As$_4$, however, it simply shows a kink instead. We thus plot the derivative of specific heat, as shown in Fig. 6(d). An obvious jump at $T_{\text{Curie}}$ can be seen in the zero-field d$C$/d$T$ data. The result suggests that the magnetic transition is of a rare third order, according to the Ehrenfest classification\cite{ehrenfest}. To the best of our knowledge, no third-order magnetic transition has been reported in a real material before. As we know, most magnetic ordering transitions are described by Landau's second-order phase transition theory. Therefore, the observation of third-order magnetic transition deserves further investigations.


\subsection{\label{subsec:level5}Discussion}

RbFe$_2$As$_2$ is known to be a 2.6 K superconductor\cite{RbFe2As2}, while EuFe$_2$As$_2$ belongs to a parent compound of FeSCs\cite{ren2008}. Here we show that, interestingly, their hybrid material RbEuFe$_4$As$_4$ turns out to be a ferromagnetic superconductor. Similar material hybridization effect in FeSCs is witnessed in Ba$_2$Ti$_2$Fe$_2$As$_4$O\cite{syl} which is actually an intergrowth of non-superconducting BaTi$_2$As$_2$O and BaFe$_2$As$_2$. Ba$_2$Ti$_2$Fe$_2$As$_4$O shows SC at 21 K owing to a charge transfer between different layers, in relation with an electron-correlation effect\cite{ding2014}. Here in RbEuFe$_4$As$_4$, the isolated ``RbFe$_2$As$_2$" block is heavily hole doped, in contrast with the undoped state in the ``EuFe$_2$As$_2$" block. Consequently, the structural hybridization leads to a charge homogenization, since there is only one equivalent Fe site. Namely, RbEuFe$_4$As$_4$ is naturally hole doped by 22.5\%. We hope that such a structural hybridization approach by design\cite{jh} may be utilized for future material explorations.

Notably, the $T_{\text{sc}}$ value of RbEuFe$_4$As$_4$ is unexpectedly high, which is nearly equal to the maximum $T_{\text{sc}}$ of 36.8 K for CsSrFe$_4$As$_4$ in the $AkAe$Fe$_4$As$_4$ series\cite{1144}. Previous studies on doped EuFe$_2$As$_2$ superconductors show that the $T_{\text{sc}}$ is always significantly lower than the Eu-free counterparts, which is ascribed to the Eu4$f-$Fe$3d$ interaction. Therefore, the unsuppressed $T_{\text{sc}}$ in RbEuFe$_4$As$_4$ suggests vanishingly small Eu4$f-$Fe$3d_{yz/zx}$ coupling. The elongation of Eu$-$Fe interatomic distance from 3.604 {\AA} in EuFe$_2$As$_2$ to 3.685 {\AA} in RbEuFe$_4$As$_4$ supports this point of view. In addition, the high value of $\mu_0$d$H_{\mathrm{c2}}$/d$T|_{T_{\mathrm{sc}}}$ ($-$5.6 T/K) as well as the large specific-heat jump (7.5 J K$^{-1}$ mol$^{-1}$) further indicates robustness of SC in RbEuFe$_4$As$_4$.


In addition to the robust SC, evidence of Eu FM is also strong and sufficient. The $M(H)$ data show an obvious magnetic hysteresis (with a coersive field of 360 Oe at 2 K) as well as a saturation magnetization (6.5 $\mu_{\text{B}}$/Eu) that corresponds to Eu$^{2+}$-spin ferromagnetic alignment. Besides, both the positive paramagnetic Curie temperature and the increase of $T_{\mathrm{Curie}}$ with magnetic field support the ferromagnetic transition scanario. Hopefully, future measurements of the anisotropic magnetic properties using single crystal samples will give more information on the magnetic state. Needless to say, investigations of the M\"{o}ssbauer spectra, x-ray resonant magnetic scattering and neutron scattering will be definitely helpful to clarity the Eu-spin orientation and other related information.

The appearance of Eu FM in RbEuFe$_4$As$_4$ can be explained in terms of modification of $J_{\mathrm{R}}^{\perp}$, since $J_{\mathrm{R}}^{\perp}$ [$\propto \mathrm{cos}(2k_{\mathrm{F}}r)/r^{3}$] \emph{oscillates} and tends to decay with $r_{\perp}$. In RbEuFe$_4$As$_4$, the interlayer Eu interatomic distance $r_{\perp}$ becomes 13.31 {\AA}, which is almost twice as the $r_{\perp}$ value in EuFe$_2$As$_2$ (6.657 {\AA}). This may change the sign of $J_{\mathrm{R}}^{\perp}$, resulting in an Eu FM. The lowered Curie temperature [about 4 K lower than that of EuFe$_2$(As$_{0.7}$P$_{0.3}$)$_2$] seems to be associated with the decay of $J_{\mathrm{R}}^{\perp}$. The weak interlayer magnetic coupling not only accounts for the relatively low Curie temperature, but also justifies the robustness of SC. Additionally, it could be related to the third-order transition observed, because of the enhanced two dimensionality which allows strong magnetic fluctuations above $T_{\mathrm{Curie}}$.



Finally, let us discuss how SC and FM compromise in RbEuFe$_4$As$_4$. There were a few theoretical proposals available on the FM+SC issue decades ago\cite{anderson,ff,lo,varma,tachiki}. The first solution is that FM is modified in the form of a ``cryptoferromagnetic" state or a multidomain structure (with domain size $d$), hence the superconducting Cooper pairs feel no net magnetization if the superconducting coherent length satisfies $\xi\gg d$\cite{anderson}. The second idea considers that superconducting Cooper pairs are ``magnetized" by the exchange fields, such that they possess non-zero momentum\cite{ff}, or equivalently, SC is modulated to be inhomogeneous in real space\cite{lo}. This scenario is often called Fulde-Ferrell-Larkin-Ovchinnikov (FFLO) state.  Alternatively, the internal spontaneous magnetization field from FM penetrates the superconductor, which induces ``spontaneous vortices" even at zero external field, hence called spontaneous vortex (SV) state\cite{varma,tachiki}. Since non-superconducting defects tend to trap spontaneous vortices, the FFLO state is believed to be preferably stabilized in a clean-limit superconductor. As for the RbEuFe$_4$As$_4$ system, the first candidate seems to be unlikely, because the estimated coherence length at zero temperature is only 1.52 nm\cite{xi_0}, which constrains the ferromagnetic domain size too much to present a long-range order. Note that the residual resistivity tends to be very small by a rough extrapolation (see Fig. 2) even for the polycrystalline sample, which means a quite long electron mean free path and, the flux-pinning effect is not severe from the FC diamagnetic signal (see Fig. 3). These facts suggest that the superconductor is probably in a clean limit at lower temperatures. We thus speculate that the anomaly at $T^{*}\sim$ 5 K could be related to the change in the way of FM+SC. There are indeed some signatures. The bifurcation of FCC and FCH curves above $T^{*}$, shown in Fig. 3, implies the transformation from FFLO to SV state. The small specific-heat jump at $T^{*}$ [Fig. 6(c)] suggests a weak phase transition. In addition, $T^{*}$ increases with field. Obviously, more investigations are needed to address this interesting issue.

\section{\label{sec:level4}Conclusion}

To summarize, we have discovered an iron-based compound RbEuFe$_4$As$_4$ in which an unprecedented coexistence of SC and FM is observed. The new material crystallizes in an intergrowth structure of body-centered RbFe$_2$As$_2$ and EuFe$_2$As$_2$, which drastically changes the physical properties. The FeAs layers are found to be hole doped by the material itself, presenting a large normal-state electronic specific-heat coefficient ($\sim$ 150 mJ K$^{-2}$ mol$^{-1}$) and a Pauli spin susceptibility ($\sim$ 0.00178 emu mol$^{-1}$). Bulk superconductivity at $T_{\mathrm{sc}}$ = 36.5 K and Eu-spin ferromagnetism at 15 K are unambiguously demonstrated. The specific-heat jump at $T_{\mathrm{sc}}$ is as high as 7.5 J K$^{-1}$ mol$^{-1}$. The Eu-spin ferromagnetism is manifested by a magnetic hysteresis with a 360-Oe coersive field at 2 K, a saturation magnetization of 6.5 $\mu_{\text{B}}$/Eu, and a saturation field of $\sim$ 1.5 kOe at low temperatures.

There are two additional novel phenomena observed in RbEuFe$_4$As$_4$. First, the Eu-spin ferromagnetic transition is of a rare third order, evidenced by the continuity in $C(T)$ and discontinuity in d$C$/d$T$ at $T_{\mathrm{Curie}}$ . Second, a weak anomaly at about 5 K possibly reflects the interplay between SC and FM. These intriguing observations together with the robust coexistence of SC and FM call for further investigations on the interesting title material.

\begin{acknowledgments}
This work was supported by the National Science Foundation of China (under grants 11474252, 90922002 and 11190023) and the Fundamental Research Funds for the Central Universities of China.
\end{acknowledgments}
%


\begin{thebibliography}{50}%
\makeatletter
\providecommand \@ifxundefined [1]{%
 \@ifx{#1\undefined}
}%
\providecommand \@ifnum [1]{%
 \ifnum #1\expandafter \@firstoftwo
 \else \expandafter \@secondoftwo
 \fi
}%
\providecommand \@ifx [1]{%
 \ifx #1\expandafter \@firstoftwo
 \else \expandafter \@secondoftwo
 \fi
}%
\providecommand \natexlab [1]{#1}%
\providecommand \enquote  [1]{``#1''}%
\providecommand \bibnamefont  [1]{#1}%
\providecommand \bibfnamefont [1]{#1}%
\providecommand \citenamefont [1]{#1}%
\providecommand \href@noop [0]{\@secondoftwo}%
\providecommand \href[0]{\begingroup \@sanitize@url \@href}%
\providecommand \@href[1]{\@@startlink{#1}\@@href}%
\providecommand \@@href[1]{\endgroup#1\@@endlink}%
\providecommand \@sanitize@url [0]{\catcode `\\12\catcode `\$12\catcode
  `\&12\catcode `\#12\catcode `\^12\catcode `\_12\catcode `\%12\relax}%
\providecommand \@@startlink[1]{}%
\providecommand \@@endlink[0]{}%
\providecommand \url  [0]{\begingroup\@sanitize@url \@url }%
\providecommand \@url [1]{\endgroup\@href {#1}{\urlprefix }}%
\providecommand \urlprefix  [0]{URL }%
\providecommand \Eprint [0]{\href }%
\providecommand \doibase [0]{http://dx.doi.org/}%
\providecommand \selectlanguage [0]{\@gobble}%
\providecommand \bibinfo  [0]{\@secondoftwo}%
\providecommand \bibfield  [0]{\@secondoftwo}%
\providecommand \translation [1]{[#1]}%
\providecommand \BibitemOpen [0]{}%
\providecommand \bibitemStop [0]{}%
\providecommand \bibitemNoStop [0]{.\EOS\space}%
\providecommand \EOS [0]{\spacefactor3000\relax}%
\providecommand \BibitemShut  [1]{\csname bibitem#1\endcsname}%
\let\auto@bib@innerbib\@empty
\bibitem [{\citenamefont {Ren}\ \emph {et~al.}(2009{\natexlab{a}})\citenamefont
  {Ren}, \citenamefont {Tao}, \citenamefont {Jiang}, \citenamefont {Feng},
  \citenamefont {Wang}, \citenamefont {Dai}, \citenamefont {Cao},\ and\
  \citenamefont {Xu}}]{ren2009}%
  \BibitemOpen
  \bibfield  {author} {\bibinfo {author} {\bibfnamefont {Z.}~\bibnamefont
  {Ren}}, \bibinfo {author} {\bibfnamefont {Q.}~\bibnamefont {Tao}}, \bibinfo
  {author} {\bibfnamefont {S.}~\bibnamefont {Jiang}}, \bibinfo {author}
  {\bibfnamefont {C.}~\bibnamefont {Feng}}, \bibinfo {author} {\bibfnamefont
  {C.}~\bibnamefont {Wang}}, \bibinfo {author} {\bibfnamefont {J.}~\bibnamefont
  {Dai}}, \bibinfo {author} {\bibfnamefont {G.}~\bibnamefont {Cao}}, \ and\
  \bibinfo {author} {\bibfnamefont {Z.}~\bibnamefont {Xu}},\ }\enquote
  {\bibinfo {title} {Superconductivity Induced by Phosphorus Doping and Its
  Coexistence with Ferromagnetism in
  ${\mathrm{EuFe}}_{2}({\mathrm{As}}_{0.7}{\mathrm{P}}_{0.3}{)}_{2}$},}\
  \href{\doibase 10.1103/PhysRevLett.102.137002} {\bibfield  {journal}
  {\bibinfo  {journal} {Phys. Rev. Lett.}\ }\textbf {\bibinfo {volume} {102}},\
  \bibinfo {pages} {137002} (\bibinfo {year} {2009}{\natexlab{a}})}\BibitemShut
  {NoStop}%
\bibitem [{\citenamefont {Nowik}\ \emph {et~al.}(2011)\citenamefont {Nowik},
  \citenamefont {Felner}, \citenamefont {Ren}, \citenamefont {Cao},\ and\
  \citenamefont {Xu}}]{felner2011}%
  \BibitemOpen
  \bibfield  {author} {\bibinfo {author} {\bibfnamefont {I.}~\bibnamefont
  {Nowik}}, \bibinfo {author} {\bibfnamefont {I.}~\bibnamefont {Felner}},
  \bibinfo {author} {\bibfnamefont {Z.}~\bibnamefont {Ren}}, \bibinfo {author}
  {\bibfnamefont {G.~H.}\ \bibnamefont {Cao}}, \ and\ \bibinfo {author}
  {\bibfnamefont {Z.~A.}\ \bibnamefont {Xu}},\ }\enquote {\bibinfo {title}
  {Coexistence of ferromagnetism and superconductivity: magnetization and
  M\"{o}ssbauer studies of
  ${\mathrm{EuFe}}_{2}({\mathrm{As}}_{1-x}{\mathrm{P}}_{x}{)}_{2}$},}\
  \href{http://stacks.iop.org/0953-8984/23/i=6/a=065701} {\bibfield  {journal}
  {\bibinfo  {journal} {J. Phys.: Condens. Matt.}\ }\textbf {\bibinfo {volume}
  {23}},\ \bibinfo {pages} {065701} (\bibinfo {year} {2011})}\BibitemShut
  {NoStop}%
\bibitem [{\citenamefont {Nandi}\ \emph
  {et~al.}(2014{\natexlab{a}})\citenamefont {Nandi}, \citenamefont {Jin},
  \citenamefont {Xiao}, \citenamefont {Su}, \citenamefont {Price},
  \citenamefont {Shukla}, \citenamefont {Strempfer}, \citenamefont {Jeevan},
  \citenamefont {Gegenwart},\ and\ \citenamefont {Br\"uckel}}]{rxs2014}%
  \BibitemOpen
  \bibfield  {author} {\bibinfo {author} {\bibfnamefont {S.}~\bibnamefont
  {Nandi}}, \bibinfo {author} {\bibfnamefont {W.~T.}\ \bibnamefont {Jin}},
  \bibinfo {author} {\bibfnamefont {Y.}~\bibnamefont {Xiao}}, \bibinfo {author}
  {\bibfnamefont {Y.}~\bibnamefont {Su}}, \bibinfo {author} {\bibfnamefont
  {S.}~\bibnamefont {Price}}, \bibinfo {author} {\bibfnamefont {D.~K.}\
  \bibnamefont {Shukla}}, \bibinfo {author} {\bibfnamefont {J.}~\bibnamefont
  {Strempfer}}, \bibinfo {author} {\bibfnamefont {H.~S.}\ \bibnamefont
  {Jeevan}}, \bibinfo {author} {\bibfnamefont {P.}~\bibnamefont {Gegenwart}}, \
  and\ \bibinfo {author} {\bibfnamefont {T.}~\bibnamefont {Br\"uckel}},\
  }\enquote {\bibinfo {title} {Coexistence of superconductivity and
  ferromagnetism in P-doped ${\text{EuFe}}_{2}{\mathrm{As}}_{2}$},}\
  \href{\doibase 10.1103/PhysRevB.89.014512} {\bibfield  {journal} {\bibinfo
  {journal} {Phys. Rev. B}\ }\textbf {\bibinfo {volume} {89}},\ \bibinfo
  {pages} {014512} (\bibinfo {year} {2014}{\natexlab{a}})}\BibitemShut
  {NoStop}%
\bibitem [{\citenamefont {Nandi}\ \emph
  {et~al.}(2014{\natexlab{b}})\citenamefont {Nandi}, \citenamefont {Jin},
  \citenamefont {Xiao}, \citenamefont {Su}, \citenamefont {Price},
  \citenamefont {Schmidt}, \citenamefont {Schmalzl}, \citenamefont {Chatterji},
  \citenamefont {Jeevan}, \citenamefont {Gegenwart},\ and\ \citenamefont
  {Br\"uckel}}]{nd2014}%
  \BibitemOpen
  \bibfield  {author} {\bibinfo {author} {\bibfnamefont {S.}~\bibnamefont
  {Nandi}}, \bibinfo {author} {\bibfnamefont {W.~T.}\ \bibnamefont {Jin}},
  \bibinfo {author} {\bibfnamefont {Y.}~\bibnamefont {Xiao}}, \bibinfo {author}
  {\bibfnamefont {Y.}~\bibnamefont {Su}}, \bibinfo {author} {\bibfnamefont
  {S.}~\bibnamefont {Price}}, \bibinfo {author} {\bibfnamefont
  {W.}~\bibnamefont {Schmidt}}, \bibinfo {author} {\bibfnamefont
  {K.}~\bibnamefont {Schmalzl}}, \bibinfo {author} {\bibfnamefont
  {T.}~\bibnamefont {Chatterji}}, \bibinfo {author} {\bibfnamefont {H.~S.}\
  \bibnamefont {Jeevan}}, \bibinfo {author} {\bibfnamefont {P.}~\bibnamefont
  {Gegenwart}}, \ and\ \bibinfo {author} {\bibfnamefont {T.}~\bibnamefont
  {Br\"uckel}},\ }\enquote {\bibinfo {title} {Magnetic structure of the ${\mathrm{Eu}}^{2+}$ moments in superconducting ${\mathrm{EuFe}}_{2}{({\mathrm{As}}_{1\ensuremath{-}x}{\mathrm{P}}_{x})}_{2}$ with $x=0.19$},}\ \href{\doibase 10.1103/PhysRevB.90.094407} {\bibfield  {journal}
  {\bibinfo  {journal} {Phys. Rev. B}\ }\textbf {\bibinfo {volume} {90}},\
  \bibinfo {pages} {094407} (\bibinfo {year} {2014}{\natexlab{b}})}\BibitemShut
  {NoStop}%
\bibitem [{\citenamefont {Ren}\ \emph {et~al.}(2008)\citenamefont {Ren},
  \citenamefont {Zhu}, \citenamefont {Jiang}, \citenamefont {Xu}, \citenamefont
  {Tao}, \citenamefont {Wang}, \citenamefont {Feng}, \citenamefont {Cao},\ and\
  \citenamefont {Xu}}]{ren2008}%
  \BibitemOpen
  \bibfield  {author} {\bibinfo {author} {\bibfnamefont {Z.}~\bibnamefont
  {Ren}}, \bibinfo {author} {\bibfnamefont {Z.}~\bibnamefont {Zhu}}, \bibinfo
  {author} {\bibfnamefont {S.}~\bibnamefont {Jiang}}, \bibinfo {author}
  {\bibfnamefont {X.}~\bibnamefont {Xu}}, \bibinfo {author} {\bibfnamefont
  {Q.}~\bibnamefont {Tao}}, \bibinfo {author} {\bibfnamefont {C.}~\bibnamefont
  {Wang}}, \bibinfo {author} {\bibfnamefont {C.}~\bibnamefont {Feng}}, \bibinfo
  {author} {\bibfnamefont {G.}~\bibnamefont {Cao}}, \ and\ \bibinfo {author}
  {\bibfnamefont {Z.}~\bibnamefont {Xu}},\ }\enquote {\bibinfo {title}
  {Antiferromagnetic transition in ${\text{EuFe}}_{2}{\text{As}}_{2}$: A
  possible parent compound for superconductors},}\ \href{\doibase
  10.1103/PhysRevB.78.052501} {\bibfield  {journal} {\bibinfo  {journal} {Phys.
  Rev. B}\ }\textbf {\bibinfo {volume} {78}},\ \bibinfo {pages} {052501}
  (\bibinfo {year} {2008})}\BibitemShut {NoStop}%
\bibitem [{\citenamefont {Jeevan}\ \emph
  {et~al.}(2008{\natexlab{a}})\citenamefont {Jeevan}, \citenamefont {Hossain},
  \citenamefont {Kasinathan}, \citenamefont {Rosner}, \citenamefont {Geibel},\
  and\ \citenamefont {Gegenwart}}]{jeevan2008}%
  \BibitemOpen
  \bibfield  {author} {\bibinfo {author} {\bibfnamefont {H.~S.}\ \bibnamefont
  {Jeevan}}, \bibinfo {author} {\bibfnamefont {Z.}~\bibnamefont {Hossain}},
  \bibinfo {author} {\bibfnamefont {D.}~\bibnamefont {Kasinathan}}, \bibinfo
  {author} {\bibfnamefont {H.}~\bibnamefont {Rosner}}, \bibinfo {author}
  {\bibfnamefont {C.}~\bibnamefont {Geibel}}, \ and\ \bibinfo {author}
  {\bibfnamefont {P.}~\bibnamefont {Gegenwart}},\ }\enquote {\bibinfo {title}
  {Electrical resistivity and specific heat of single-crystalline
  ${\text{EuFe}}_{2}{\text{As}}_{2}$: A magnetic homologue of
  ${\text{SrFe}}_{2}{\text{As}}_{2}$},}\ \href{\doibase
  10.1103/PhysRevB.78.052502} {\bibfield  {journal} {\bibinfo  {journal} {Phys.
  Rev. B}\ }\textbf {\bibinfo {volume} {78}},\ \bibinfo {pages} {052502}
  (\bibinfo {year} {2008}{\natexlab{a}})}\BibitemShut {NoStop}%
\bibitem [{\citenamefont {Jiang}\ \emph
  {et~al.}(2009{\natexlab{a}})\citenamefont {Jiang}, \citenamefont {Luo},
  \citenamefont {Ren}, \citenamefont {Zhu}, \citenamefont {Wang}, \citenamefont
  {Xu}, \citenamefont {Tao}, \citenamefont {Cao},\ and\ \citenamefont
  {Xu}}]{js-njp}%
  \BibitemOpen
  \bibfield  {author} {\bibinfo {author} {\bibfnamefont {S.}~\bibnamefont
  {Jiang}}, \bibinfo {author} {\bibfnamefont {Y.}~\bibnamefont {Luo}}, \bibinfo
  {author} {\bibfnamefont {Z.}~\bibnamefont {Ren}}, \bibinfo {author}
  {\bibfnamefont {Z.}~\bibnamefont {Zhu}}, \bibinfo {author} {\bibfnamefont
  {C.}~\bibnamefont {Wang}}, \bibinfo {author} {\bibfnamefont {X.}~\bibnamefont
  {Xu}}, \bibinfo {author} {\bibfnamefont {Q.}~\bibnamefont {Tao}}, \bibinfo
  {author} {\bibfnamefont {G.}~\bibnamefont {Cao}}, \ and\ \bibinfo {author}
  {\bibfnamefont {Z.}~\bibnamefont {Xu}},\ }\enquote {\bibinfo {title}
  {Metamagnetic transition in ${\text{EuFe}}_{2}{\text{As}}_{2}$ single
  crystals},}\ \href{http://stacks.iop.org/1367-2630/11/i=2/a=025007}
  {\bibfield  {journal} {\bibinfo  {journal} {New J. Phys.}\ }\textbf {\bibinfo
  {volume} {11}},\ \bibinfo {pages} {025007} (\bibinfo {year}
  {2009}{\natexlab{a}})}\BibitemShut {NoStop}%
\bibitem [{\citenamefont {Herrero-Mart\'{i}n}\ \emph
  {et~al.}(2009)\citenamefont {Herrero-Mart\'{i}n}, \citenamefont {Scagnoli},
  \citenamefont {Mazzoli}, \citenamefont {Su}, \citenamefont {Mittal},
  \citenamefont {Xiao}, \citenamefont {Brueckel}, \citenamefont {Kumar},
  \citenamefont {Dhar}, \citenamefont {Thamizhavel},\ and\ \citenamefont
  {Paolasini}}]{rxs2009}%
  \BibitemOpen
  \bibfield  {author} {\bibinfo {author} {\bibfnamefont {J.}~\bibnamefont
  {Herrero-Mart\'{i}n}}, \bibinfo {author} {\bibfnamefont {V.}~\bibnamefont
  {Scagnoli}}, \bibinfo {author} {\bibfnamefont {C.}~\bibnamefont {Mazzoli}},
  \bibinfo {author} {\bibfnamefont {Y.}~\bibnamefont {Su}}, \bibinfo {author}
  {\bibfnamefont {R.}~\bibnamefont {Mittal}}, \bibinfo {author} {\bibfnamefont
  {Y.}~\bibnamefont {Xiao}}, \bibinfo {author} {\bibfnamefont {T.}~\bibnamefont
  {Brueckel}}, \bibinfo {author} {\bibfnamefont {N.}~\bibnamefont {Kumar}},
  \bibinfo {author} {\bibfnamefont {S.~K.}\ \bibnamefont {Dhar}}, \bibinfo
  {author} {\bibfnamefont {A.}~\bibnamefont {Thamizhavel}}, \ and\ \bibinfo
  {author} {\bibfnamefont {L.}~\bibnamefont {Paolasini}},\ }\enquote {\bibinfo
  {title} {Magnetic structure of ${\text{EuFe}}_{2}{\text{As}}_{2}$ as
  determined by resonant x-ray scattering},}\ \href{\doibase
  10.1103/PhysRevB.80.134411} {\bibfield  {journal} {\bibinfo  {journal} {Phys.
  Rev. B}\ }\textbf {\bibinfo {volume} {80}},\ \bibinfo {pages} {134411}
  (\bibinfo {year} {2009})}\BibitemShut {NoStop}%
\bibitem [{\citenamefont {Xiao}\ \emph {et~al.}(2009)\citenamefont {Xiao},
  \citenamefont {Su}, \citenamefont {Meven}, \citenamefont {Mittal},
  \citenamefont {Kumar}, \citenamefont {Chatterji}, \citenamefont {Price},
  \citenamefont {Persson}, \citenamefont {Kumar}, \citenamefont {Dhar},
  \citenamefont {Thamizhavel},\ and\ \citenamefont {Brueckel}}]{nd2009}%
  \BibitemOpen
  \bibfield  {author} {\bibinfo {author} {\bibfnamefont {Y.}~\bibnamefont
  {Xiao}}, \bibinfo {author} {\bibfnamefont {Y.}~\bibnamefont {Su}}, \bibinfo
  {author} {\bibfnamefont {M.}~\bibnamefont {Meven}}, \bibinfo {author}
  {\bibfnamefont {R.}~\bibnamefont {Mittal}}, \bibinfo {author} {\bibfnamefont
  {C.~M.~N.}\ \bibnamefont {Kumar}}, \bibinfo {author} {\bibfnamefont
  {T.}~\bibnamefont {Chatterji}}, \bibinfo {author} {\bibfnamefont
  {S.}~\bibnamefont {Price}}, \bibinfo {author} {\bibfnamefont
  {J.}~\bibnamefont {Persson}}, \bibinfo {author} {\bibfnamefont
  {N.}~\bibnamefont {Kumar}}, \bibinfo {author} {\bibfnamefont {S.~K.}\
  \bibnamefont {Dhar}}, \bibinfo {author} {\bibfnamefont {A.}~\bibnamefont
  {Thamizhavel}}, \ and\ \bibinfo {author} {\bibfnamefont {T.}~\bibnamefont
  {Brueckel}},\ }\enquote {\bibinfo {title} {Magnetic structure of
  ${\text{EuFe}}_{2}{\text{As}}_{2}$ determined by single-crystal neutron
  diffraction},}\ \href{\doibase 10.1103/PhysRevB.80.174424} {\bibfield
  {journal} {\bibinfo  {journal} {Phys. Rev. B}\ }\textbf {\bibinfo {volume}
  {80}},\ \bibinfo {pages} {174424} (\bibinfo {year} {2009})}\BibitemShut
  {NoStop}%
\bibitem [{\citenamefont {Cao}\ \emph {et~al.}(2011)\citenamefont {Cao},
  \citenamefont {Xu}, \citenamefont {Ren}, \citenamefont {Jiang}, \citenamefont
  {Feng},\ and\ \citenamefont {Xu}}]{cao2011}%
  \BibitemOpen
  \bibfield  {author} {\bibinfo {author} {\bibfnamefont {G.}~\bibnamefont
  {Cao}}, \bibinfo {author} {\bibfnamefont {S.}~\bibnamefont {Xu}}, \bibinfo
  {author} {\bibfnamefont {Z.}~\bibnamefont {Ren}}, \bibinfo {author}
  {\bibfnamefont {S.}~\bibnamefont {Jiang}}, \bibinfo {author} {\bibfnamefont
  {C.}~\bibnamefont {Feng}}, \ and\ \bibinfo {author} {\bibfnamefont
  {Z.}~\bibnamefont {Xu}},\ }\enquote {\bibinfo {title} {Superconductivity and
  ferromagnetism in EuFe$_2$(As$_{1-x}$P$_x$)$_2$},}\
  \href{http://stacks.iop.org/0953-8984/23/i=46/a=464204} {\bibfield  {journal}
  {\bibinfo  {journal} {J. Phys.: Condens. Matt.}\ }\textbf {\bibinfo {volume}
  {23}},\ \bibinfo {pages} {464204} (\bibinfo {year} {2011})}\BibitemShut
  {NoStop}%
\bibitem [{\citenamefont {Jeevan}\ \emph {et~al.}(2011)\citenamefont {Jeevan},
  \citenamefont {Kasinathan}, \citenamefont {Rosner},\ and\ \citenamefont
  {Gegenwart}}]{jeevan2011}%
  \BibitemOpen
  \bibfield  {author} {\bibinfo {author} {\bibfnamefont {H.~S.}\ \bibnamefont
  {Jeevan}}, \bibinfo {author} {\bibfnamefont {D.}~\bibnamefont {Kasinathan}},
  \bibinfo {author} {\bibfnamefont {H.}~\bibnamefont {Rosner}}, \ and\ \bibinfo
  {author} {\bibfnamefont {P.}~\bibnamefont {Gegenwart}},\ }\enquote {\bibinfo
  {title} {Interplay of antiferromagnetism, ferromagnetism, and
  superconductivity in EuFe${}_{2}$(As${}_{1-x}$P${}_{x}$)${}_{2}$ single
  crystals},}\ \href{\doibase 10.1103/PhysRevB.83.054511} {\bibfield  {journal}
  {\bibinfo  {journal} {Phys. Rev. B}\ }\textbf {\bibinfo {volume} {83}},\
  \bibinfo {pages} {054511} (\bibinfo {year} {2011})}\BibitemShut {NoStop}%
\bibitem [{\citenamefont {Jiang}\ \emph
  {et~al.}(2009{\natexlab{b}})\citenamefont {Jiang}, \citenamefont {Xing},
  \citenamefont {Xuan}, \citenamefont {Ren}, \citenamefont {Wang},
  \citenamefont {Xu},\ and\ \citenamefont {Cao}}]{jiang2009}%
  \BibitemOpen
  \bibfield  {author} {\bibinfo {author} {\bibfnamefont {S.}~\bibnamefont
  {Jiang}}, \bibinfo {author} {\bibfnamefont {H.}~\bibnamefont {Xing}},
  \bibinfo {author} {\bibfnamefont {G.}~\bibnamefont {Xuan}}, \bibinfo {author}
  {\bibfnamefont {Z.}~\bibnamefont {Ren}}, \bibinfo {author} {\bibfnamefont
  {C.}~\bibnamefont {Wang}}, \bibinfo {author} {\bibfnamefont {Z.}~\bibnamefont
  {Xu}}, \ and\ \bibinfo {author} {\bibfnamefont {G.}~\bibnamefont {Cao}},\
  }\enquote {\bibinfo {title} {Superconductivity and local-moment magnetism in
  $\text{Eu}{({\text{Fe}}_{0.89}{\text{Co}}_{0.11})}_{2}{\text{As}}_{2}$},}\
  \href{\doibase 10.1103/PhysRevB.80.184514} {\bibfield  {journal} {\bibinfo
  {journal} {Phys. Rev. B}\ }\textbf {\bibinfo {volume} {80}},\ \bibinfo
  {pages} {184514} (\bibinfo {year} {2009}{\natexlab{b}})}\BibitemShut
  {NoStop}%
\bibitem [{\citenamefont {Jiao}\ \emph {et~al.}(2011)\citenamefont {Jiao},
  \citenamefont {Tao}, \citenamefont {Bao}, \citenamefont {Sun}, \citenamefont
  {Feng}, \citenamefont {Xu}, \citenamefont {Nowik}, \citenamefont {Felner},\
  and\ \citenamefont {Cao}}]{jiao2011}%
  \BibitemOpen
  \bibfield  {author} {\bibinfo {author} {\bibfnamefont {W.-H.}\ \bibnamefont
  {Jiao}}, \bibinfo {author} {\bibfnamefont {Q.}~\bibnamefont {Tao}}, \bibinfo
  {author} {\bibfnamefont {J.-K.}\ \bibnamefont {Bao}}, \bibinfo {author}
  {\bibfnamefont {Y.-L.}\ \bibnamefont {Sun}}, \bibinfo {author} {\bibfnamefont
  {C.-M.}\ \bibnamefont {Feng}}, \bibinfo {author} {\bibfnamefont {Z.-A.}\
  \bibnamefont {Xu}}, \bibinfo {author} {\bibfnamefont {I.}~\bibnamefont
  {Nowik}}, \bibinfo {author} {\bibfnamefont {I.}~\bibnamefont {Felner}}, \
  and\ \bibinfo {author} {\bibfnamefont {G.-H.}\ \bibnamefont {Cao}},\
  }\enquote {\bibinfo {title} {Anisotropic superconductivity in
  Eu(Fe$_{0.75}$Ru$_{0.25}$)$_2$As$_2$ ferromagnetic superconductor},}\
  \href{http://stacks.iop.org/0295-5075/95/i=6/a=67007} {\bibfield  {journal}
  {\bibinfo  {journal} {EPL (Europhysics Letters)}\ }\textbf {\bibinfo {volume}
  {95}},\ \bibinfo {pages} {67007} (\bibinfo {year} {2011})}\BibitemShut
  {NoStop}%
\bibitem [{\citenamefont {Jiao}\ \emph {et~al.}(2013)\citenamefont {Jiao},
  \citenamefont {Zhai}, \citenamefont {Bao}, \citenamefont {Luo}, \citenamefont
  {Tao}, \citenamefont {Feng}, \citenamefont {Xu},\ and\ \citenamefont
  {Cao}}]{jiao2013}%
  \BibitemOpen
  \bibfield  {author} {\bibinfo {author} {\bibfnamefont {W.-H.}\ \bibnamefont
  {Jiao}}, \bibinfo {author} {\bibfnamefont {H.-F.}\ \bibnamefont {Zhai}},
  \bibinfo {author} {\bibfnamefont {J.-K.}\ \bibnamefont {Bao}}, \bibinfo
  {author} {\bibfnamefont {Y.-K.}\ \bibnamefont {Luo}}, \bibinfo {author}
  {\bibfnamefont {Q.}~\bibnamefont {Tao}}, \bibinfo {author} {\bibfnamefont
  {C.-M.}\ \bibnamefont {Feng}}, \bibinfo {author} {\bibfnamefont {Z.-A.}\
  \bibnamefont {Xu}}, \ and\ \bibinfo {author} {\bibfnamefont {G.-H.}\
  \bibnamefont {Cao}},\ }\enquote {\bibinfo {title} {Anomalous critical fields
  and the absence of Meissner state in Eu(Fe$_{0.88}$Ir$_{0.12}$)$_2$As$_2$
  crystals},}\ \href{http://stacks.iop.org/1367-2630/15/i=11/a=113002}
  {\bibfield  {journal} {\bibinfo  {journal} {New J. Phys.}\ }\textbf {\bibinfo
  {volume} {15}},\ \bibinfo {pages} {113002} (\bibinfo {year}
  {2013})}\BibitemShut {NoStop}%
\bibitem [{\citenamefont {Jin}\ \emph {et~al.}(2013)\citenamefont {Jin},
  \citenamefont {Nandi}, \citenamefont {Xiao}, \citenamefont {Su},
  \citenamefont {Zaharko}, \citenamefont {Guguchia}, \citenamefont {Bukowski},
  \citenamefont {Price}, \citenamefont {Jiao}, \citenamefont {Cao},\ and\
  \citenamefont {Br\"uckel}}]{jin-Co}%
  \BibitemOpen
  \bibfield  {author} {\bibinfo {author} {\bibfnamefont {W.~T.}\ \bibnamefont
  {Jin}}, \bibinfo {author} {\bibfnamefont {S.}~\bibnamefont {Nandi}}, \bibinfo
  {author} {\bibfnamefont {Y.}~\bibnamefont {Xiao}}, \bibinfo {author}
  {\bibfnamefont {Y.}~\bibnamefont {Su}}, \bibinfo {author} {\bibfnamefont
  {O.}~\bibnamefont {Zaharko}}, \bibinfo {author} {\bibfnamefont
  {Z.}~\bibnamefont {Guguchia}}, \bibinfo {author} {\bibfnamefont
  {Z.}~\bibnamefont {Bukowski}}, \bibinfo {author} {\bibfnamefont
  {S.}~\bibnamefont {Price}}, \bibinfo {author} {\bibfnamefont {W.~H.}\
  \bibnamefont {Jiao}}, \bibinfo {author} {\bibfnamefont {G.~H.}\ \bibnamefont
  {Cao}}, \ and\ \bibinfo {author} {\bibfnamefont {T.}~\bibnamefont
  {Br\"uckel}},\ }\enquote {\bibinfo {title} {Magnetic structure of
  superconducting Eu(Fe${}_{0.82}$Co${}_{0.18}$)${}_{2}$As${}_{2}$ as revealed
  by single-crystal neutron diffraction},}\ \href{\doibase
  10.1103/PhysRevB.88.214516} {\bibfield  {journal} {\bibinfo  {journal} {Phys.
  Rev. B}\ }\textbf {\bibinfo {volume} {88}},\ \bibinfo {pages} {214516}
  (\bibinfo {year} {2013})}\BibitemShut {NoStop}%
\bibitem [{\citenamefont {Jin}\ \emph {et~al.}(2015)\citenamefont {Jin},
  \citenamefont {Li}, \citenamefont {Su}, \citenamefont {Nandi}, \citenamefont
  {Xiao}, \citenamefont {Jiao}, \citenamefont {Meven}, \citenamefont {Sazonov},
  \citenamefont {Feng}, \citenamefont {Chen}, \citenamefont {Ting},
  \citenamefont {Cao},\ and\ \citenamefont {Br\"uckel}}]{jin-Ir}%
  \BibitemOpen
  \bibfield  {author} {\bibinfo {author} {\bibfnamefont {W.~T.}\ \bibnamefont
  {Jin}}, \bibinfo {author} {\bibfnamefont {W.}~\bibnamefont {Li}}, \bibinfo
  {author} {\bibfnamefont {Y.}~\bibnamefont {Su}}, \bibinfo {author}
  {\bibfnamefont {S.}~\bibnamefont {Nandi}}, \bibinfo {author} {\bibfnamefont
  {Y.}~\bibnamefont {Xiao}}, \bibinfo {author} {\bibfnamefont {W.~H.}\
  \bibnamefont {Jiao}}, \bibinfo {author} {\bibfnamefont {M.}~\bibnamefont
  {Meven}}, \bibinfo {author} {\bibfnamefont {A.~P.}\ \bibnamefont {Sazonov}},
  \bibinfo {author} {\bibfnamefont {E.}~\bibnamefont {Feng}}, \bibinfo {author}
  {\bibfnamefont {Y.}~\bibnamefont {Chen}}, \bibinfo {author} {\bibfnamefont
  {C.~S.}\ \bibnamefont {Ting}}, \bibinfo {author} {\bibfnamefont {G.~H.}\
  \bibnamefont {Cao}}, \ and\ \bibinfo {author} {\bibfnamefont
  {T.}~\bibnamefont {Br\"uckel}},\ }\enquote {\bibinfo {title} {Magnetic ground
  state of superconducting
  $\mathrm{Eu}(\mathrm{Fe}{}_{0.88}\mathrm{Ir}{}_{0.12}){}_{2}\mathrm{As}{}_{2}$:
  A combined neutron diffraction and first-principles calculation study},}\
  \href{\doibase 10.1103/PhysRevB.91.064506} {\bibfield  {journal} {\bibinfo
  {journal} {Phys. Rev. B}\ }\textbf {\bibinfo {volume} {91}},\ \bibinfo
  {pages} {064506} (\bibinfo {year} {2015})}\BibitemShut {NoStop}%
\bibitem [{\citenamefont {Tokiwa}\ \emph {et~al.}(2012)\citenamefont {Tokiwa},
  \citenamefont {H\"ubner}, \citenamefont {Beck}, \citenamefont {Jeevan},\ and\
  \citenamefont {Gegenwart}}]{tokiwa}%
  \BibitemOpen
  \bibfield  {author} {\bibinfo {author} {\bibfnamefont {Y.}~\bibnamefont
  {Tokiwa}}, \bibinfo {author} {\bibfnamefont {S.-H.}\ \bibnamefont
  {H\"ubner}}, \bibinfo {author} {\bibfnamefont {O.}~\bibnamefont {Beck}},
  \bibinfo {author} {\bibfnamefont {H.~S.}\ \bibnamefont {Jeevan}}, \ and\
  \bibinfo {author} {\bibfnamefont {P.}~\bibnamefont {Gegenwart}},\ }\enquote
  {\bibinfo {title} {Unique phase diagram with narrow superconducting dome in
  EuFe${}_{2}$(As${}_{1\ensuremath{-}x}$P${}_{x}$)${}_{2}$ due to Eu${}^{2+}$
  local magnetic moments},}\ \href{\doibase 10.1103/PhysRevB.86.220505}
  {\bibfield  {journal} {\bibinfo  {journal} {Phys. Rev. B}\ }\textbf {\bibinfo
  {volume} {86}},\ \bibinfo {pages} {220505} (\bibinfo {year}
  {2012})}\BibitemShut {NoStop}%
\bibitem [{\citenamefont {Zapf}\ \emph {et~al.}(2011)\citenamefont {Zapf},
  \citenamefont {Wu}, \citenamefont {Bogani}, \citenamefont {Jeevan},
  \citenamefont {Gegenwart},\ and\ \citenamefont {Dressel}}]{zapf2011}%
  \BibitemOpen
  \bibfield  {author} {\bibinfo {author} {\bibfnamefont {S.}~\bibnamefont
  {Zapf}}, \bibinfo {author} {\bibfnamefont {D.}~\bibnamefont {Wu}}, \bibinfo
  {author} {\bibfnamefont {L.}~\bibnamefont {Bogani}}, \bibinfo {author}
  {\bibfnamefont {H.~S.}\ \bibnamefont {Jeevan}}, \bibinfo {author}
  {\bibfnamefont {P.}~\bibnamefont {Gegenwart}}, \ and\ \bibinfo {author}
  {\bibfnamefont {M.}~\bibnamefont {Dressel}},\ }\enquote {\bibinfo {title}
  {Varying Eu${}^{2+}$ magnetic order by chemical pressure in
  EuFe${}_{2}$(As${}_{1\ensuremath{-}x}$P${}_{x}$)${}_{2}$},}\ \href{\doibase
  10.1103/PhysRevB.84.140503} {\bibfield  {journal} {\bibinfo  {journal} {Phys.
  Rev. B}\ }\textbf {\bibinfo {volume} {84}},\ \bibinfo {pages} {140503}
  (\bibinfo {year} {2011})}\BibitemShut {NoStop}%
\bibitem [{\citenamefont {Zapf}\ \emph {et~al.}(2013)\citenamefont {Zapf},
  \citenamefont {Jeevan}, \citenamefont {Ivek}, \citenamefont {Pfister},
  \citenamefont {Klingert}, \citenamefont {Jiang}, \citenamefont {Wu},
  \citenamefont {Gegenwart}, \citenamefont {Kremer},\ and\ \citenamefont
  {Dressel}}]{zapf2013}%
  \BibitemOpen
  \bibfield  {author} {\bibinfo {author} {\bibfnamefont {S.}~\bibnamefont
  {Zapf}}, \bibinfo {author} {\bibfnamefont {H.~S.}\ \bibnamefont {Jeevan}},
  \bibinfo {author} {\bibfnamefont {T.}~\bibnamefont {Ivek}}, \bibinfo {author}
  {\bibfnamefont {F.}~\bibnamefont {Pfister}}, \bibinfo {author} {\bibfnamefont
  {F.}~\bibnamefont {Klingert}}, \bibinfo {author} {\bibfnamefont
  {S.}~\bibnamefont {Jiang}}, \bibinfo {author} {\bibfnamefont
  {D.}~\bibnamefont {Wu}}, \bibinfo {author} {\bibfnamefont {P.}~\bibnamefont
  {Gegenwart}}, \bibinfo {author} {\bibfnamefont {R.~K.}\ \bibnamefont
  {Kremer}}, \ and\ \bibinfo {author} {\bibfnamefont {M.}~\bibnamefont
  {Dressel}},\ }\enquote {\bibinfo {title}
  {${\mathrm{EuFe}}_{2}({\mathrm{As}}_{1-x}{\mathrm{P}}_{x}{)}_{2}$: Reentrant
  Spin Glass and Superconductivity},}\ \href{\doibase
  10.1103/PhysRevLett.110.237002} {\bibfield  {journal} {\bibinfo  {journal}
  {Phys. Rev. Lett.}\ }\textbf {\bibinfo {volume} {110}},\ \bibinfo {pages}
  {237002} (\bibinfo {year} {2013})}\BibitemShut {NoStop}%
\bibitem [{\citenamefont {B\l{}achowski}\ \emph {et~al.}(2011)\citenamefont
  {B\l{}achowski}, \citenamefont {Ruebenbauer}, \citenamefont
  {\ifmmode~\dot{Z}\else \.{Z}\fi{}ukrowski}, \citenamefont {Bukowski},
  \citenamefont {Rogacki}, \citenamefont {Moll},\ and\ \citenamefont
  {Karpinski}}]{blachowski}%
  \BibitemOpen
  \bibfield  {author} {\bibinfo {author} {\bibfnamefont {A.}~\bibnamefont
  {B\l{}achowski}}, \bibinfo {author} {\bibfnamefont {K.}~\bibnamefont
  {Ruebenbauer}}, \bibinfo {author} {\bibfnamefont {J.}~\bibnamefont
  {\ifmmode~\dot{Z}\else \.{Z}\fi{}ukrowski}}, \bibinfo {author} {\bibfnamefont
  {Z.}~\bibnamefont {Bukowski}}, \bibinfo {author} {\bibfnamefont
  {K.}~\bibnamefont {Rogacki}}, \bibinfo {author} {\bibfnamefont {P.~J.~W.}\
  \bibnamefont {Moll}}, \ and\ \bibinfo {author} {\bibfnamefont
  {J.}~\bibnamefont {Karpinski}},\ }\enquote {\bibinfo {title}
  {Interplay between magnetism and superconductivity in EuFe${}_{2\ensuremath{-}x}$Co${}_{x}$As${}_{2}$ studied by ${}^{57}$Fe and ${}^{151}$Eu M\"{o}ssbauer spectroscopy},}\ \href {\doibase 10.1103/PhysRevB.84.174503}
  {\bibfield  {journal} {\bibinfo  {journal} {Phys. Rev. B}\ }\textbf {\bibinfo
  {volume} {84}},\ \bibinfo {pages} {174503} (\bibinfo {year}
  {2011})}\BibitemShut {NoStop}%
\bibitem [{\citenamefont {Ginzburg}(1957)}]{ginzburg}%
  \BibitemOpen
  \bibfield  {author} {\bibinfo {author} {\bibfnamefont {V.~L.}\ \bibnamefont
  {Ginzburg}},\ }\enquote {\bibinfo {title} {Ferromagnetic Superconductors},}\
  \href@noop {} {\bibfield  {journal} {\bibinfo  {journal} {Soviet Physics
  JETP-USSR}\ }\textbf {\bibinfo {volume} {4}},\ \bibinfo {pages} {153}
  (\bibinfo {year} {1957})}\BibitemShut {NoStop}%
\bibitem [{\citenamefont {Bulaevskii}\ \emph {et~al.}(1985)\citenamefont
  {Bulaevskii}, \citenamefont {Buzdin}, \citenamefont {Kulic},\ and\
  \citenamefont {Panjukov}}]{rev1985}%
  \BibitemOpen
  \bibfield  {author} {\bibinfo {author} {\bibfnamefont {L.~N.}\ \bibnamefont
  {Bulaevskii}}, \bibinfo {author} {\bibfnamefont {A.~I.}\ \bibnamefont
  {Buzdin}}, \bibinfo {author} {\bibfnamefont {M.~L.}\ \bibnamefont {Kulic}}, \
  and\ \bibinfo {author} {\bibfnamefont {S.~V.}\ \bibnamefont {Panjukov}},\
  }\enquote {\bibinfo {title} {Coexistence of Superconductivity and Magnetism -
  Theoretical Predictions and Experimental Results},}\ \href{\doibase
  10.1080/00018738500101741} {\bibfield  {journal} {\bibinfo  {journal} {Adv.
  Phys.}\ }\textbf {\bibinfo {volume} {34}},\ \bibinfo {pages} {175} (\bibinfo
  {year} {1985})}\BibitemShut {NoStop}%
\bibitem [{\citenamefont {Wolowiec}\ \emph {et~al.}(2015)\citenamefont
  {Wolowiec}, \citenamefont {White},\ and\ \citenamefont {Maple}}]{maple}%
  \BibitemOpen
  \bibfield  {author} {\bibinfo {author} {\bibfnamefont {C.}~\bibnamefont
  {Wolowiec}}, \bibinfo {author} {\bibfnamefont {B.}~\bibnamefont {White}}, \
  and\ \bibinfo {author} {\bibfnamefont {M.}~\bibnamefont {Maple}},\ }\enquote
  {\bibinfo {title} {Conventional magnetic superconductors},}\ \href{\doibase
  http://dx.doi.org/10.1016/j.physc.2015.02.050} {\bibfield  {journal}
  {\bibinfo  {journal} {Physica C}\ }\textbf {\bibinfo {volume} {514}},\
  \bibinfo {pages} {113 } (\bibinfo {year} {2015})}\BibitemShut {NoStop}%
\bibitem [{fms()}]{fmsc}%
  \BibitemOpen
  \href@noop {} {}\bibinfo {note} {Note that spin-triplet superconductivity may
  coexist with itinerant ferromagnetism in some U-containing materials. The
  related infomation can be seen in a recent review article [A. D. Huxley,
  \emph{Ferromagnetic superconductors}, Physica C \textbf{514}, 368-377
  (2015)]}\BibitemShut {NoStop}%
\bibitem [{\citenamefont {Cao}\ \emph {et~al.}(2012)\citenamefont {Cao},
  \citenamefont {Jiao}, \citenamefont {Luo}, \citenamefont {Ren}, \citenamefont
  {Jiang},\ and\ \citenamefont {Xu}}]{cao2012}%
  \BibitemOpen
  \bibfield  {author} {\bibinfo {author} {\bibfnamefont {G.-H.}\ \bibnamefont
  {Cao}}, \bibinfo {author} {\bibfnamefont {W.-H.}\ \bibnamefont {Jiao}},
  \bibinfo {author} {\bibfnamefont {Y.-K.}\ \bibnamefont {Luo}}, \bibinfo
  {author} {\bibfnamefont {Z.}~\bibnamefont {Ren}}, \bibinfo {author}
  {\bibfnamefont {S.}~\bibnamefont {Jiang}}, \ and\ \bibinfo {author}
  {\bibfnamefont {Z.-A.}\ \bibnamefont {Xu}},\ }\enquote {\bibinfo {title}
  {Coexistence of superconductivity and ferromagnetism in iron pnictides},}\
  \href{http://stacks.iop.org/1742-6596/391/i=1/a=012123} {\bibfield  {journal}
  {\bibinfo  {journal} {J. Phys.: Conf. Ser.}\ }\textbf {\bibinfo {volume}
  {391}},\ \bibinfo {pages} {012123} (\bibinfo {year} {2012})}\BibitemShut
  {NoStop}%
\bibitem [{\citenamefont {Johnston}(2010)}]{johnston}%
  \BibitemOpen
  \bibfield  {author} {\bibinfo {author} {\bibfnamefont {D.~C.}\ \bibnamefont
  {Johnston}},\ }\enquote {\bibinfo {title} {The puzzle of high temperature
  superconductivity in layered iron pnictides and chalcogenides},}\
  \href{http://dx.doi.org/10.1080/00018732.2010.513480} {\bibfield  {journal}
  {\bibinfo  {journal} {Adv. Phys.}\ }\textbf {\bibinfo {volume} {59}},\
  \bibinfo {pages} {803} (\bibinfo {year} {2010})}\BibitemShut {NoStop}%
\bibitem [{\citenamefont {Stewart}(2011)}]{stewart}%
  \BibitemOpen
  \bibfield  {author} {\bibinfo {author} {\bibfnamefont {G.~R.}\ \bibnamefont
  {Stewart}},\ }\enquote {\bibinfo {title} {Superconductivity in iron
  compounds},}\ \href{\doibase 10.1103/RevModPhys.83.1589} {\bibfield
  {journal} {\bibinfo  {journal} {Rev. Mod. Phys.}\ }\textbf {\bibinfo {volume}
  {83}},\ \bibinfo {pages} {1589} (\bibinfo {year} {2011})}\BibitemShut
  {NoStop}%
\bibitem [{\citenamefont {Raghu}\ \emph {et~al.}(2008)\citenamefont {Raghu},
  \citenamefont {Qi}, \citenamefont {Liu}, \citenamefont {Scalapino},\ and\
  \citenamefont {Zhang}}]{zhangsc}%
  \BibitemOpen
  \bibfield  {author} {\bibinfo {author} {\bibfnamefont {S.}~\bibnamefont
  {Raghu}}, \bibinfo {author} {\bibfnamefont {X.-L.}\ \bibnamefont {Qi}},
  \bibinfo {author} {\bibfnamefont {C.-X.}\ \bibnamefont {Liu}}, \bibinfo
  {author} {\bibfnamefont {D.~J.}\ \bibnamefont {Scalapino}}, \ and\ \bibinfo
  {author} {\bibfnamefont {S.-C.}\ \bibnamefont {Zhang}},\ }\enquote {\bibinfo
  {title} {Minimal two-band model of the superconducting iron oxypnictides},}\
  \href{\doibase 10.1103/PhysRevB.77.220503} {\bibfield  {journal} {\bibinfo
  {journal} {Phys. Rev. B}\ }\textbf {\bibinfo {volume} {77}},\ \bibinfo
  {pages} {220503} (\bibinfo {year} {2008})}\BibitemShut {NoStop}%
\bibitem [{\citenamefont {Ren}\ \emph {et~al.}(2009{\natexlab{b}})\citenamefont
  {Ren}, \citenamefont {Lin}, \citenamefont {Tao}, \citenamefont {Jiang},
  \citenamefont {Zhu}, \citenamefont {Wang}, \citenamefont {Cao},\ and\
  \citenamefont {Xu}}]{ren-Ni}%
  \BibitemOpen
  \bibfield  {author} {\bibinfo {author} {\bibfnamefont {Z.}~\bibnamefont
  {Ren}}, \bibinfo {author} {\bibfnamefont {X.}~\bibnamefont {Lin}}, \bibinfo
  {author} {\bibfnamefont {Q.}~\bibnamefont {Tao}}, \bibinfo {author}
  {\bibfnamefont {S.}~\bibnamefont {Jiang}}, \bibinfo {author} {\bibfnamefont
  {Z.}~\bibnamefont {Zhu}}, \bibinfo {author} {\bibfnamefont {C.}~\bibnamefont
  {Wang}}, \bibinfo {author} {\bibfnamefont {G.}~\bibnamefont {Cao}}, \ and\
  \bibinfo {author} {\bibfnamefont {Z.}~\bibnamefont {Xu}},\ }\enquote
  {\bibinfo {title} {Suppression of spin-density-wave transition and emergence
  of ferromagnetic ordering of ${\text{Eu}}^{2+}$ moments in
  ${\text{EuFe}}_{2\ensuremath{-}x}{\text{Ni}}_{x}{\text{As}}_{2}$},}\
  \href{\doibase 10.1103/PhysRevB.79.094426} {\bibfield  {journal} {\bibinfo
  {journal} {Phys. Rev. B}\ }\textbf {\bibinfo {volume} {79}},\ \bibinfo
  {pages} {094426} (\bibinfo {year} {2009}{\natexlab{b}})}\BibitemShut
  {NoStop}%
\bibitem [{\citenamefont {Singh}\ and\ \citenamefont {Du}(2008)}]{singh}%
  \BibitemOpen
  \bibfield  {author} {\bibinfo {author} {\bibfnamefont {D.~J.}\ \bibnamefont
  {Singh}}\ and\ \bibinfo {author} {\bibfnamefont {M.-H.}\ \bibnamefont {Du}},\
  }\enquote {\bibinfo {title} {Density Functional Study of
  ${\mathrm{LaFeAsO}}_{1-x}{\mathrm{F}}_{x}$: A Low Carrier Density
  Superconductor Near Itinerant Magnetism},}\ \href{\doibase
  10.1103/PhysRevLett.100.237003} {\bibfield  {journal} {\bibinfo  {journal}
  {Phys. Rev. Lett.}\ }\textbf {\bibinfo {volume} {100}},\ \bibinfo {pages}
  {237003} (\bibinfo {year} {2008})}\BibitemShut {NoStop}%
\bibitem [{\citenamefont {Li}\ \emph {et~al.}(2012)\citenamefont {Li},
  \citenamefont {Zhu}, \citenamefont {Chen},\ and\ \citenamefont {Ting}}]{LDA}%
  \BibitemOpen
  \bibfield  {author} {\bibinfo {author} {\bibfnamefont {W.}~\bibnamefont
  {Li}}, \bibinfo {author} {\bibfnamefont {J.-X.}\ \bibnamefont {Zhu}},
  \bibinfo {author} {\bibfnamefont {Y.}~\bibnamefont {Chen}}, \ and\ \bibinfo
  {author} {\bibfnamefont {C.~S.}\ \bibnamefont {Ting}},\ }\enquote {\bibinfo
  {title} {First-principles calculations of the electronic structure of
  iron-pnictide EuFe${}_{2}$(As,P)${}_{2}$ superconductors: Evidence for
  antiferromagnetic spin order},}\ \href{\doibase 10.1103/PhysRevB.86.155119}
  {\bibfield  {journal} {\bibinfo  {journal} {Phys. Rev. B}\ }\textbf {\bibinfo
  {volume} {86}},\ \bibinfo {pages} {155119} (\bibinfo {year}
  {2012})}\BibitemShut {NoStop}%
\bibitem [{\citenamefont {Jiang}\ \emph {et~al.}(2013)\citenamefont {Jiang},
  \citenamefont {Sun}, \citenamefont {Xu},\ and\ \citenamefont {Cao}}]{jh}%
  \BibitemOpen
  \bibfield  {author} {\bibinfo {author} {\bibfnamefont {H.}~\bibnamefont
  {Jiang}}, \bibinfo {author} {\bibfnamefont {Y.-L.}\ \bibnamefont {Sun}},
  \bibinfo {author} {\bibfnamefont {Z.-A.}\ \bibnamefont {Xu}}, \ and\ \bibinfo
  {author} {\bibfnamefont {G.-H.}\ \bibnamefont {Cao}},\ }\enquote {\bibinfo
  {title} {Crystal chemistry and structural design of iron-based
  superconductors},}\ \href{http://stacks.iop.org/1674-1056/22/i=8/a=087410}
  {\bibfield  {journal} {\bibinfo  {journal} {Chin. Phys. B}\ }\textbf
  {\bibinfo {volume} {22}},\ \bibinfo {pages} {087410} (\bibinfo {year}
  {2013})}\BibitemShut {NoStop}%
\bibitem [{\citenamefont {Iyo}\ \emph {et~al.}(2016)\citenamefont {Iyo},
  \citenamefont {Kawashima}, \citenamefont {Kinjo}, \citenamefont {Nishio},
  \citenamefont {Ishida}, \citenamefont {Fujihisa}, \citenamefont {Gotoh},
  \citenamefont {Kihou}, \citenamefont {Eisaki},\ and\ \citenamefont
  {Yoshida}}]{1144}%
  \BibitemOpen
  \bibfield  {author} {\bibinfo {author} {\bibfnamefont {A.}~\bibnamefont
  {Iyo}}, \bibinfo {author} {\bibfnamefont {K.}~\bibnamefont {Kawashima}},
  \bibinfo {author} {\bibfnamefont {T.}~\bibnamefont {Kinjo}}, \bibinfo
  {author} {\bibfnamefont {T.}~\bibnamefont {Nishio}}, \bibinfo {author}
  {\bibfnamefont {S.}~\bibnamefont {Ishida}}, \bibinfo {author} {\bibfnamefont
  {H.}~\bibnamefont {Fujihisa}}, \bibinfo {author} {\bibfnamefont
  {Y.}~\bibnamefont {Gotoh}}, \bibinfo {author} {\bibfnamefont
  {K.}~\bibnamefont {Kihou}}, \bibinfo {author} {\bibfnamefont
  {H.}~\bibnamefont {Eisaki}}, \ and\ \bibinfo {author} {\bibfnamefont
  {Y.}~\bibnamefont {Yoshida}},\ }\enquote {\bibinfo {title}
  {New-Structure-Type Fe-Based Superconductors: CaAFe$_4$As$_4$ (A = K, Rb, Cs)
  and SrAFe$_4$As$_4$ (A = Rb, Cs)},}\ \href{\doibase 10.1021/jacs.5b12571}
  {\bibfield  {journal} {\bibinfo  {journal} {J. Am. Chem. Soc.}\ }\textbf
  {\bibinfo {volume} {138}},\ \bibinfo {pages} {3410} (\bibinfo {year}
  {2016})}\BibitemShut {NoStop}%
\bibitem [{\citenamefont {Izumi}\ and\ \citenamefont
  {Momma}(2007)}]{rietan-fp2}%
  \BibitemOpen
  \bibfield  {author} {\bibinfo {author} {\bibfnamefont {F.}~\bibnamefont
  {Izumi}}\ and\ \bibinfo {author} {\bibfnamefont {K.}~\bibnamefont {Momma}},\
  }\enquote {\bibinfo {title} {Three-dimensional visualization in powder
  diffraction},}\ in\ \href{<Go to ISI>://WOS:000249626200003} {\emph {\bibinfo
  {booktitle} {Applied Crystallography XX}}},\ \bibinfo {series} {Solid State
  Phenomena}, Vol.\ \bibinfo {volume} {130},\ \bibinfo {editor} {edited by\
  \bibinfo {editor} {\bibfnamefont {D.}~\bibnamefont {Stroz}}\ and\ \bibinfo
  {editor} {\bibfnamefont {M.}~\bibnamefont {Karolus}}}\ (\bibinfo {year}
  {2007})\ pp.\ \bibinfo {pages} {15--20}\BibitemShut {NoStop}%
\bibitem [{\citenamefont {Sample}\ \emph {et~al.}(1987)\citenamefont {Sample},
  \citenamefont {Bruno}, \citenamefont {Sample},\ and\ \citenamefont
  {Sichel}}]{sample1987}%
  \BibitemOpen
  \bibfield  {author} {\bibinfo {author} {\bibfnamefont {H.~H.}\ \bibnamefont
  {Sample}}, \bibinfo {author} {\bibfnamefont {W.~J.}\ \bibnamefont {Bruno}},
  \bibinfo {author} {\bibfnamefont {S.~B.}\ \bibnamefont {Sample}}, \ and\
  \bibinfo {author} {\bibfnamefont {E.~K.}\ \bibnamefont {Sichel}},\ }\enquote
  {\bibinfo {title} {Reverse-field reciprocity for conducting specimens in
  magnetic fields},}\ \href@noop {} {\bibfield  {journal} {\bibinfo  {journal}
  {J. Appl. Phys.}\ }\textbf {\bibinfo {volume} {61}} (\bibinfo {year}
  {1987})}\BibitemShut {NoStop}%
\bibitem [{\citenamefont {Wenz}\ and\ \citenamefont {Schuster}(1984)}]{Rb122}%
  \BibitemOpen
  \bibfield  {author} {\bibinfo {author} {\bibfnamefont {P.}~\bibnamefont
  {Wenz}}\ and\ \bibinfo {author} {\bibfnamefont {H.~U.}\ \bibnamefont
  {Schuster}},\ }\enquote {\bibinfo {title} {New Ternary Intermetallic Phases
  of Potasium and Rubidium with 8B-Elements and 5B-Elements},}\ \href{<Go to
  ISI>://WOS:A1984AAV2700033} {\bibfield  {journal} {\bibinfo  {journal} {Z.
  Nat. Sect. B}\ }\textbf {\bibinfo {volume} {39}},\ \bibinfo {pages} {1816}
  (\bibinfo {year} {1984})}\BibitemShut {NoStop}%
\bibitem [{\citenamefont {Rotter}\ \emph {et~al.}(2008)\citenamefont {Rotter},
  \citenamefont {Tegel},\ and\ \citenamefont {Johrendt}}]{BaK}%
  \BibitemOpen
  \bibfield  {author} {\bibinfo {author} {\bibfnamefont {M.}~\bibnamefont
  {Rotter}}, \bibinfo {author} {\bibfnamefont {M.}~\bibnamefont {Tegel}}, \
  and\ \bibinfo {author} {\bibfnamefont {D.}~\bibnamefont {Johrendt}},\
  }\enquote {\bibinfo {title} {Superconductivity at 38 K in the Iron Arsenide
  $({\mathrm{Ba}}_{1-x}{\mathrm{K}}_{x}){\mathrm{Fe}}_{2}{\mathrm{As}}_{2}$},}\
  \href{\doibase 10.1103/PhysRevLett.101.107006} {\bibfield  {journal}
  {\bibinfo  {journal} {Phys. Rev. Lett.}\ }\textbf {\bibinfo {volume} {101}},\
  \bibinfo {pages} {107006} (\bibinfo {year} {2008})}\BibitemShut {NoStop}%
\bibitem [{\citenamefont {Shen}\ \emph {et~al.}(2011)\citenamefont {Shen},
  \citenamefont {Yang}, \citenamefont {Wang}, \citenamefont {Han},
  \citenamefont {Zeng}, \citenamefont {Shan}, \citenamefont {Ren},\ and\
  \citenamefont {Wen}}]{whh-transport}%
  \BibitemOpen
  \bibfield  {author} {\bibinfo {author} {\bibfnamefont {B.}~\bibnamefont
  {Shen}}, \bibinfo {author} {\bibfnamefont {H.}~\bibnamefont {Yang}}, \bibinfo
  {author} {\bibfnamefont {Z.-S.}\ \bibnamefont {Wang}}, \bibinfo {author}
  {\bibfnamefont {F.}~\bibnamefont {Han}}, \bibinfo {author} {\bibfnamefont
  {B.}~\bibnamefont {Zeng}}, \bibinfo {author} {\bibfnamefont {L.}~\bibnamefont
  {Shan}}, \bibinfo {author} {\bibfnamefont {C.}~\bibnamefont {Ren}}, \ and\
  \bibinfo {author} {\bibfnamefont {H.-H.}\ \bibnamefont {Wen}},\ }\enquote
  {\bibinfo {title} {Transport properties and asymmetric scattering in
  Ba${}_{1\ensuremath{-}x}$K${}_{x}$Fe${}_{2}$As${}_{2}$ single crystals},}\
  \href{\doibase 10.1103/PhysRevB.84.184512} {\bibfield  {journal} {\bibinfo
  {journal} {Phys. Rev. B}\ }\textbf {\bibinfo {volume} {84}},\ \bibinfo
  {pages} {184512} (\bibinfo {year} {2011})}\BibitemShut {NoStop}%
\bibitem [{\citenamefont {Jeevan}\ \emph
  {et~al.}(2008{\natexlab{b}})\citenamefont {Jeevan}, \citenamefont {Hossain},
  \citenamefont {Kasinathan}, \citenamefont {Rosner}, \citenamefont {Geibel},\
  and\ \citenamefont {Gegenwart}}]{EuK}%
  \BibitemOpen
  \bibfield  {author} {\bibinfo {author} {\bibfnamefont {H.~S.}\ \bibnamefont
  {Jeevan}}, \bibinfo {author} {\bibfnamefont {Z.}~\bibnamefont {Hossain}},
  \bibinfo {author} {\bibfnamefont {D.}~\bibnamefont {Kasinathan}}, \bibinfo
  {author} {\bibfnamefont {H.}~\bibnamefont {Rosner}}, \bibinfo {author}
  {\bibfnamefont {C.}~\bibnamefont {Geibel}}, \ and\ \bibinfo {author}
  {\bibfnamefont {P.}~\bibnamefont {Gegenwart}},\ }\enquote {\bibinfo {title}
  {High-temperature superconductivity in
  ${\text{Eu}}_{0.5}{\text{K}}_{0.5}{\text{Fe}}_{2}{\text{As}}_{2}$},}\
  \href{\doibase 10.1103/PhysRevB.78.092406} {\bibfield  {journal} {\bibinfo
  {journal} {Phys. Rev. B}\ }\textbf {\bibinfo {volume} {78}},\ \bibinfo
  {pages} {092406} (\bibinfo {year} {2008}{\natexlab{b}})}\BibitemShut
  {NoStop}%
\bibitem [{\citenamefont {Fertig}\ \emph {et~al.}(1977)\citenamefont {Fertig},
  \citenamefont {Johnston}, \citenamefont {DeLong}, \citenamefont {McCallum},
  \citenamefont {Maple},\ and\ \citenamefont {Matthias}}]{ErRh4B4}%
  \BibitemOpen
  \bibfield  {author} {\bibinfo {author} {\bibfnamefont {W.~A.}\ \bibnamefont
  {Fertig}}, \bibinfo {author} {\bibfnamefont {D.~C.}\ \bibnamefont
  {Johnston}}, \bibinfo {author} {\bibfnamefont {L.~E.}\ \bibnamefont
  {DeLong}}, \bibinfo {author} {\bibfnamefont {R.~W.}\ \bibnamefont
  {McCallum}}, \bibinfo {author} {\bibfnamefont {M.~B.}\ \bibnamefont {Maple}},
  \ and\ \bibinfo {author} {\bibfnamefont {B.~T.}\ \bibnamefont {Matthias}},\
  }\enquote {\bibinfo {title} {Destruction of Superconductivity at the Onset of
  Long-Range Magnetic Order in the Compound
  Er${\mathrm{Rh}}_{4}$${\mathrm{B}}_{4}$},}\ \href{\doibase
  10.1103/PhysRevLett.38.987} {\bibfield  {journal} {\bibinfo  {journal} {Phys.
  Rev. Lett.}\ }\textbf {\bibinfo {volume} {38}},\ \bibinfo {pages} {987}
  (\bibinfo {year} {1977})}\BibitemShut {NoStop}%
\bibitem [{\citenamefont {Mu}\ \emph {et~al.}(2009)\citenamefont {Mu},
  \citenamefont {Luo}, \citenamefont {Wang}, \citenamefont {Shan},
  \citenamefont {Ren},\ and\ \citenamefont {Wen}}]{whh2009}%
  \BibitemOpen
  \bibfield  {author} {\bibinfo {author} {\bibfnamefont {G.}~\bibnamefont
  {Mu}}, \bibinfo {author} {\bibfnamefont {H.}~\bibnamefont {Luo}}, \bibinfo
  {author} {\bibfnamefont {Z.}~\bibnamefont {Wang}}, \bibinfo {author}
  {\bibfnamefont {L.}~\bibnamefont {Shan}}, \bibinfo {author} {\bibfnamefont
  {C.}~\bibnamefont {Ren}}, \ and\ \bibinfo {author} {\bibfnamefont {H.-H.}\
  \bibnamefont {Wen}},\ }\enquote {\bibinfo {title} {Low temperature specific
  heat of the hole-doped
  ${\text{Ba}}_{0.6}{\text{K}}_{0.4}{\text{Fe}}_{2}{\text{As}}_{2}$ single
  crystals},}\ \href{\doibase 10.1103/PhysRevB.79.174501} {\bibfield  {journal}
  {\bibinfo  {journal} {Phys. Rev. B}\ }\textbf {\bibinfo {volume} {79}},\
  \bibinfo {pages} {174501} (\bibinfo {year} {2009})}\BibitemShut {NoStop}%
\bibitem [{\citenamefont {Jaeger}(1998)}]{ehrenfest}%
  \BibitemOpen
  \bibfield  {author} {\bibinfo {author} {\bibfnamefont {G.}~\bibnamefont
  {Jaeger}},\ }\enquote {\bibinfo {title} {The Ehrenfest Classification of
  Phase Transitions: Introduction and Evolution},}\ \href{\doibase
  10.1007/s004070050021} {\bibfield  {journal} {\bibinfo  {journal} {Arch.
  Hist. Exact Sci.}\ }\textbf {\bibinfo {volume} {53}},\ \bibinfo {pages} {51}
  (\bibinfo {year} {1998})}\BibitemShut {NoStop}%
\bibitem [{\citenamefont {Bukowski}\ \emph {et~al.}(2010)\citenamefont
  {Bukowski}, \citenamefont {Weyeneth}, \citenamefont {Puzniak}, \citenamefont
  {Karpinski},\ and\ \citenamefont {Batlogg}}]{RbFe2As2}%
  \BibitemOpen
  \bibfield  {author} {\bibinfo {author} {\bibfnamefont {Z.}~\bibnamefont
  {Bukowski}}, \bibinfo {author} {\bibfnamefont {S.}~\bibnamefont {Weyeneth}},
  \bibinfo {author} {\bibfnamefont {R.}~\bibnamefont {Puzniak}}, \bibinfo
  {author} {\bibfnamefont {J.}~\bibnamefont {Karpinski}}, \ and\ \bibinfo
  {author} {\bibfnamefont {B.}~\bibnamefont {Batlogg}},\ }\enquote {\bibinfo
  {title} {Bulk superconductivity at 2.6 K in undoped RbFe$_2$As$_2$},}\
  \href{\doibase http://dx.doi.org/10.1016/j.physc.2009.11.103} {\bibfield
  {journal} {\bibinfo  {journal} {Physica C}\ }\textbf {\bibinfo {volume} {470,
  Suppl. 1}},\ \bibinfo {pages} {S328 } (\bibinfo {year} {2010})}\BibitemShut
  {NoStop}%
\bibitem [{\citenamefont {Sun}\ \emph {et~al.}(2012)\citenamefont {Sun},
  \citenamefont {Jiang}, \citenamefont {Zhai}, \citenamefont {Bao},
  \citenamefont {Jiao}, \citenamefont {Tao}, \citenamefont {Shen},
  \citenamefont {Zeng}, \citenamefont {Xu},\ and\ \citenamefont {Cao}}]{syl}%
  \BibitemOpen
  \bibfield  {author} {\bibinfo {author} {\bibfnamefont {Y.-L.}\ \bibnamefont
  {Sun}}, \bibinfo {author} {\bibfnamefont {H.}~\bibnamefont {Jiang}}, \bibinfo
  {author} {\bibfnamefont {H.-F.}\ \bibnamefont {Zhai}}, \bibinfo {author}
  {\bibfnamefont {J.-K.}\ \bibnamefont {Bao}}, \bibinfo {author} {\bibfnamefont
  {W.-H.}\ \bibnamefont {Jiao}}, \bibinfo {author} {\bibfnamefont
  {Q.}~\bibnamefont {Tao}}, \bibinfo {author} {\bibfnamefont {C.-Y.}\
  \bibnamefont {Shen}}, \bibinfo {author} {\bibfnamefont {Y.-W.}\ \bibnamefont
  {Zeng}}, \bibinfo {author} {\bibfnamefont {Z.-A.}\ \bibnamefont {Xu}}, \ and\
  \bibinfo {author} {\bibfnamefont {G.-H.}\ \bibnamefont {Cao}},\ }\enquote
  {\bibinfo {title} {Ba$_2$Ti$_2$Fe$_2$As$_4$O: A New Superconductor Containing
  Fe$_2$As$_2$ Layers and Ti$_2$O Sheets},}\ \href{\doibase 10.1021/ja304315e}
  {\bibfield  {journal} {\bibinfo  {journal} {J. Am. Chem. Soc.}\ }\textbf
  {\bibinfo {volume} {134}},\ \bibinfo {pages} {12893} (\bibinfo {year}
  {2012})}\BibitemShut {NoStop}%
\bibitem [{\citenamefont {Ma}\ \emph {et~al.}(2014)\citenamefont {Ma},
  \citenamefont {van Roekeghem}, \citenamefont {Richard}, \citenamefont {Liu},
  \citenamefont {Miao}, \citenamefont {Zeng}, \citenamefont {Xu}, \citenamefont
  {Shi}, \citenamefont {Cao}, \citenamefont {He}, \citenamefont {Chen},
  \citenamefont {Sun}, \citenamefont {Cao}, \citenamefont {Wang}, \citenamefont
  {Biermann}, \citenamefont {Qian},\ and\ \citenamefont {Ding}}]{ding2014}%
  \BibitemOpen
  \bibfield  {author} {\bibinfo {author} {\bibfnamefont {J.-Z.}\ \bibnamefont
  {Ma}}, \bibinfo {author} {\bibfnamefont {A.}~\bibnamefont {van Roekeghem}},
  \bibinfo {author} {\bibfnamefont {P.}~\bibnamefont {Richard}}, \bibinfo
  {author} {\bibfnamefont {Z.-H.}\ \bibnamefont {Liu}}, \bibinfo {author}
  {\bibfnamefont {H.}~\bibnamefont {Miao}}, \bibinfo {author} {\bibfnamefont
  {L.-K.}\ \bibnamefont {Zeng}}, \bibinfo {author} {\bibfnamefont
  {N.}~\bibnamefont {Xu}}, \bibinfo {author} {\bibfnamefont {M.}~\bibnamefont
  {Shi}}, \bibinfo {author} {\bibfnamefont {C.}~\bibnamefont {Cao}}, \bibinfo
  {author} {\bibfnamefont {J.-B.}\ \bibnamefont {He}}, \bibinfo {author}
  {\bibfnamefont {G.-F.}\ \bibnamefont {Chen}}, \bibinfo {author}
  {\bibfnamefont {Y.-L.}\ \bibnamefont {Sun}}, \bibinfo {author} {\bibfnamefont
  {G.-H.}\ \bibnamefont {Cao}}, \bibinfo {author} {\bibfnamefont {S.-C.}\
  \bibnamefont {Wang}}, \bibinfo {author} {\bibfnamefont {S.}~\bibnamefont
  {Biermann}}, \bibinfo {author} {\bibfnamefont {T.}~\bibnamefont {Qian}}, \
  and\ \bibinfo {author} {\bibfnamefont {H.}~\bibnamefont {Ding}},\ }\enquote
  {\bibinfo {title} {Correlation-Induced Self-Doping in the Iron-Pnictide
  Superconductor
  ${\mathrm{Ba}}_{2}{\mathrm{Ti}}_{2}{\mathrm{Fe}}_{2}{\mathrm{As}}_{4}\mathrm{O}$},}\
  \href{\doibase 10.1103/PhysRevLett.113.266407} {\bibfield  {journal}
  {\bibinfo  {journal} {Phys. Rev. Lett.}\ }\textbf {\bibinfo {volume} {113}},\
  \bibinfo {pages} {266407} (\bibinfo {year} {2014})}\BibitemShut {NoStop}%
\bibitem [{\citenamefont {Anderson}\ and\ \citenamefont
  {Suhl}(1959)}]{anderson}%
  \BibitemOpen
  \bibfield  {author} {\bibinfo {author} {\bibfnamefont {P.~W.}\ \bibnamefont
  {Anderson}}\ and\ \bibinfo {author} {\bibfnamefont {H.}~\bibnamefont
  {Suhl}},\ }\enquote {\bibinfo {title} {Spin Alignment in the Superconducting
  State},}\ \href{\doibase 10.1103/PhysRev.116.898} {\bibfield  {journal}
  {\bibinfo  {journal} {Phys. Rev.}\ }\textbf {\bibinfo {volume} {116}},\
  \bibinfo {pages} {898} (\bibinfo {year} {1959})}\BibitemShut {NoStop}%
\bibitem [{\citenamefont {Fulde}\ and\ \citenamefont {Ferrell}(1964)}]{ff}%
  \BibitemOpen
  \bibfield  {author} {\bibinfo {author} {\bibfnamefont {P.}~\bibnamefont
  {Fulde}}\ and\ \bibinfo {author} {\bibfnamefont {R.~A.}\ \bibnamefont
  {Ferrell}},\ }\enquote {\bibinfo {title} {Superconductivity in a Strong
  Spin-Exchange Field},}\ \href{\doibase 10.1103/PhysRev.135.A550} {\bibfield
  {journal} {\bibinfo  {journal} {Phys. Rev.}\ }\textbf {\bibinfo {volume}
  {135}},\ \bibinfo {pages} {A550} (\bibinfo {year} {1964})}\BibitemShut
  {NoStop}%
\bibitem [{\citenamefont {Larkin}\ and\ \citenamefont
  {Ovchinni.Yn}(1965)}]{lo}%
  \BibitemOpen
  \bibfield  {author} {\bibinfo {author} {\bibfnamefont {A.~I.}\ \bibnamefont
  {Larkin}}\ and\ \bibinfo {author} {\bibnamefont {Ovchinni.Yn}},\ }\enquote
  {\bibinfo {title} {Inhomogeneous State of Superconductors},}\ \href{<Go to
  ISI>://WOS:A19656224900049} {\bibfield  {journal} {\bibinfo  {journal} {Sov.
  Phys. JETP-USSR}\ }\textbf {\bibinfo {volume} {20}},\ \bibinfo {pages} {762}
  (\bibinfo {year} {1965})}\BibitemShut {NoStop}%
\bibitem [{\citenamefont {Greenside}\ \emph {et~al.}(1981)\citenamefont
  {Greenside}, \citenamefont {Blount},\ and\ \citenamefont {Varma}}]{varma}%
  \BibitemOpen
  \bibfield  {author} {\bibinfo {author} {\bibfnamefont {H.~S.}\ \bibnamefont
  {Greenside}}, \bibinfo {author} {\bibfnamefont {E.~I.}\ \bibnamefont
  {Blount}}, \ and\ \bibinfo {author} {\bibfnamefont {C.~M.}\ \bibnamefont
  {Varma}},\ }\enquote {\bibinfo {title} {Possible Coexisting Superconducting
  and Magnetic States},}\ \href{\doibase 10.1103/PhysRevLett.46.49} {\bibfield
  {journal} {\bibinfo  {journal} {Phys. Rev. Lett.}\ }\textbf {\bibinfo
  {volume} {46}},\ \bibinfo {pages} {49} (\bibinfo {year} {1981})}\BibitemShut
  {NoStop}%
\bibitem [{\citenamefont {Tachiki}\ \emph {et~al.}(1980)\citenamefont
  {Tachiki}, \citenamefont {Matsumoto}, \citenamefont {Koyama},\ and\
  \citenamefont {Umezawa}}]{tachiki}%
  \BibitemOpen
  \bibfield  {author} {\bibinfo {author} {\bibfnamefont {M.}~\bibnamefont
  {Tachiki}}, \bibinfo {author} {\bibfnamefont {H.}~\bibnamefont {Matsumoto}},
  \bibinfo {author} {\bibfnamefont {T.}~\bibnamefont {Koyama}}, \ and\ \bibinfo
  {author} {\bibfnamefont {H.}~\bibnamefont {Umezawa}},\ }\enquote {\bibinfo
  {title} {Self-induced vortices in magnetic superconductors},}\ \href{\doibase
  http://dx.doi.org/10.1016/0038-1098(80)90620-1} {\bibfield  {journal}
  {\bibinfo  {journal} {Solid State Commun.}\ }\textbf {\bibinfo {volume}
  {34}},\ \bibinfo {pages} {19 } (\bibinfo {year} {1980})}\BibitemShut
  {NoStop}%
\bibitem [{xi_()}]{xi_0}%
  \BibitemOpen
  \href@noop {} {\enquote {\bibinfo {title} {The coherence length is calculated
  by $\xi(0)=\phi_{0}/[2\pi H_{\mathrm{c2}}^{\mathrm{orb}}(0)]$, where
  $H_{\mathrm{c2}}^{\mathrm{orb}}(0)$ is obtained by the slope of
  $H_{\mathrm{c2}}(T)$ ($-$5.6 T/K) using the werthamer-helfand-hohenberg
  formula   $H_{\mathrm{c2}}^{\mathrm{orb}}(0)=-0.693T_{\mathrm{sc}}(\mu_0$d$H_{\mathrm{c2}}$/d$T|_{T_{\mathrm{sc}}})=$
  141.6 T.}}\ }\BibitemShut {NoStop}%
\end{thebibliography}%
\end{document}